\documentclass[journal]{IEEEtran}

\usepackage{cite}
\usepackage{amsmath, amsfonts, amssymb, commath}
\usepackage{bm}
\usepackage{color}
\usepackage{graphicx}
\usepackage{caption}
\usepackage{subcaption}

\usepackage[ruled]{algorithm2e}
\SetKwProg{Init}{Initialize:}{}{}
\usepackage{enumerate}

\newcommand{\CN}{\mathcal{CN}}
\newcommand{\zerobf}{\bm{0}}
\newcommand{\der}{\mathrm{d}}

\newcommand{\abf}{\bm{a}}

\newcommand{\hbf}{\bm{h}}
\newcommand{\nbf}{\bm{n}}

\newcommand{\qbf}{\bm{q}}

\newcommand{\ybf}{\bm{y}}

\newcommand{\xbfhat}{{\bm{\hat{x}}}}
\newcommand{\bhat}{\hat{b}}
\newcommand{\chat}{\hat{c}}
\newcommand{\dhat}{\hat{d}}
\newcommand{\ehat}{\hat{e}}
\newcommand{\xhat}{\hat{x}}
\newcommand{\what}{\hat{w}}
\newcommand{\ghat}{\hat{g}}
\newcommand{\shat}{\hat{s}}
\newcommand{\ohat}{\hat{o}}
\newcommand{\phat}{\hat{p}}

\newcommand{\zhat}{\hat{z}}
\newcommand{\ghatit}{\hat{\textit{g}}}
\newcommand{\alphahat}{\hat{\alpha}}
\newcommand{\gammahat}{\hat{\gamma}}
\newcommand{\betahat}{\hat{\beta}}

\newcommand{\Abf}{\bm{A}}

\newcommand{\Gbf}{\bm{G}}
\newcommand{\Hbf}{\bm{H}}

\newcommand{\Mbf}{\bm{M}}
\newcommand{\Nbf}{\bm{N}}
\newcommand{\Qbf}{\bm{Q}}
\newcommand{\Sbf}{\bm{S}}

\newcommand{\Wbf}{\bm{W}}
\newcommand{\Ybf}{\bm{Y}}
\newcommand{\Xbf}{\bm{X}}
\newcommand{\Zbf}{\bm{Z}}
\newcommand{\Abfhat}{\bm{\hat{A}}}
\newcommand{\Xbfhat}{{\bm{\hat{X}}}}

\newcommand{\Phibf}{\bm{\Phi}}

\newcommand{\Cbb}{\mathbb{C}}

\newcommand{\tran}{^\mathrm{T}}
\newcommand{\herm}{^\mathrm{H}}

\def\tp1{t +  1}

\newcommand{\diag}[1]{\mathrm{diag}\{{#1}\}}

\begin{document}

\title{\huge{Reconfigurable Intelligent Surface for Massive Connectivity}}
\author{\IEEEauthorblockN{Shuhao Xia,~\IEEEmembership{Student Member,~IEEE,}
        Yuanming Shi,~\IEEEmembership{Senior Member,~IEEE,}
        Yong Zhou,~\IEEEmembership{Member,~IEEE,}
        and Xiaojun Yuan,~\IEEEmembership{Senior Member,~IEEE,}}
    \thanks{
		S. Xia, Y. Shi, and Y. Zhou are with the School of Information Science and Technology, ShanghaiTech University, Shanghai, 201210, China (E-mail: \{xiashh, shiym, zhouyong\}@shanghaitech.edu.cn).}
	\thanks{
		X. Yuan is with the Center for Intelligent Networking and Communications, the National Laboratory of Science and Technology on Communications, the University of Electronic Science and Technology of China, Chengdu, 611731, China (E-mail: xjyuan@uestc.edu.cn).}
 }

\maketitle

\begin{abstract}
With the rapid development of Internet of Things (IoT), massive machine-type communication has become a promising application scenario, where a large number of devices transmit sporadically to a base station (BS). Reconfigurable intelligent surface (RIS) has been recently proposed as an innovative new technology to achieve energy efficiency and coverage enhancement by establishing favorable signal propagation environments, thereby improving
data transmission in massive connectivity. Nevertheless, the BS needs to detect active devices and estimate channels to support data transmission in RIS-assisted massive access systems, which yields unique challenges. This paper shall consider an RIS-assisted uplink IoT network and aims to solve the \emph{RIS-related activity detection and channel estimation} problem, where the BS detects the active devices and estimates the separated channels of the RIS-to-device link and the RIS-to-BS link. Due to limited scattering between the RIS and the BS, we model the RIS-to-BS channel as a sparse channel. As a result, by simultaneously exploiting both the sparsity of sporadic transmission in massive connectivity and the RIS-to-BS channels, we formulate the RIS-related activity detection and channel estimation problem as a sparse matrix factorization problem. Furthermore, we develop an approximate message passing (AMP) based algorithm to solve the problem based on Bayesian inference framework and reduce the computational complexity by approximating the algorithm with the central limit theorem and Taylor series arguments. Finally, extensive numerical experiments are conducted to verify the effectiveness and improvements of the proposed algorithm.
\end{abstract}

\begin{IEEEkeywords}
Device activity detection, channel estimation, reconfigurable intelligent
surface, approximate message passing, Internet of Things (IoT), massive machine-type communications (mMTC).
\end{IEEEkeywords}

\section{Introduction}
With the popularity of the Internet of Things (IoT), massive connectivity, also known as massive machine-type communication (mMTC), has been regarded as one of the three typical use cases of the fifth-generation (5G) wireless networks \cite{8808168}, along with enhanced mobile broadband (eMBB) and ultra-reliable low-latency communication (uRLLC). A main challenge of mMTC is to support sporadic short-packet communications between the base station (BS) and a massive number of IoT devices \cite{7565189}. Grant-free random access that has the potential to significantly reduce the signaling overhead and access latency has recently attracted considerable attention \cite{8454392}, where multiple active IoT devices directly transmit their unique pilot sequences together with the data without the need of obtaining the grant from the BS. With grant-free random access, the BS generally needs to perform activity detection, channel estimation, and data decoding. 
Various advanced activity detection and channel estimation algorithms have been proposed in the literature. By exploiting the sporadic nature of the device activity patterns, several compressed sensing (CS)-based approaches were proposed to solve the joint activity detection and channel estimation problem, which is formulated as a sparse signal recovery problem \cite{8444464, 8536396, 8264818}. Taking into account the short-packet transmission, a covariance-based approach was proposed to jointly detect active devices and decode the data for massive connectivity \cite{8761672, 9053200}. In addition, the authors in \cite{9140386} developed a joint design of activity detection, channel estimation, and data decoding, and proposed an approximated message passing (AMP) based algorithm for a trilinear model.

In massive connectivity, the IoT devices are usually located in a service \emph{dead zone}, where the line-of-sight communication may not be available. Consequently, the signals received from the IoT devices are usually very weak, which makes the accurate device detection a challenging task for the BS. On the other hand, poor channel conditions in the dead zones also reduce the link reliability for data transmissions between the active devices and the BS \cite{9205230}. To overcome these challenges, a variety of massive access techniques have been recently proposed \cite{7827017, 8734871, 5595728, 6251827, 8794743, 6294504, 2019arXiv191206040H, 6824752}. In particular, more BSs can be deployed to shorten the communication distance to enhance the coverage and capacity \cite{7827017, 8734871}, which inevitably leads to higher energy consumption and deployment costs. Furthermore, the use of millimeter wave (mmWave) and terahertz (THz) frequencies was proposed to enhance the capacity of the IoT networks \cite{6294504, 2019arXiv191206040H, 6824752}. To mitigate the severe propagation loss over distance, mmWave/THz communications are generally combined with massive multiple-input-multiple-output (MIMO) \cite{5595728, 6251827, 8794743}, which requires increased hardware and energy cost as well as signal processing complexity. To address the aforementioned issues and limitations, it is exigent to develop novel technologies to support massive connectivity with low cost, low complexity, and high energy efficiency.


Reconfigurable intelligent surface (RIS) has recently emerged as a promising technology for enhancing the spectral efficiency and energy efficiency in various wireless communication systems \cite{2020arXiv200100364Y, 9066923, 9110912, 8683663}. To be specific, RIS is a man-made surface equipped with a large number of passive and programmable reflecting elements integrated with a smart controller \cite{8319526}. RIS plays a similar role as a large-scale antenna array through performing spatial beamforming, but with lower hardware and energy cost \cite{8811733, 8930608}. By optimizing the phase shifts based on the instantaneous channel state information (CSI), the signal propagation between the BS and the IoT devices can be smartly reconfigured to improve the quality of the data transmission. Moreover, the deployment of RIS creates more non-line-of-sight links between the BS and the devices in dead zones via bypassing the obstacle between them \cite{9140329, 9122596}. For cell-edge IoT devices, the deployment of RIS not only helps improve the desired signal power at cell-edge users, but also facilitates the suppression of co-channel interference from neighboring cells \cite{8910627}. Hence, RIS has been regarded as a promising technique to achieve coverage and capacity enhancement for massive connectivity \cite{9205230}.

In RIS-assisted IoT systems, the aforementioned performance gains depend on the availability of CSI at the BS. However, acquiring accurate CSI in RIS-assisted wireless networks is much more challenging than that in conventional wireless networks \cite{9053695, 9133142, 9127834}. 
Specifically, the RIS is typically not equipped with any radio frequency (RF) chains, and thus lacks the ability to transmit and receive signals. Consequently, the RIS is not able to perform activity detection and channel estimation between the RIS and active devices. Furthermore, the design of passive beamforming at the RIS requires the separate CSI of the RIS-to-device link and the RIS-to-BS link \cite{8811733, 8930608, 9110912}. As a result, the BS is responsible for the tasks of detecting device activity and decoupling the cascaded channels of RIS-to-device and RIS-to-BS links. In order to separately estimate the channels for RIS-assisted communication systems, various strategies have been recently proposed \cite{9104305, 8879620, 9133156, 2019arXiv191203619C, 2019arXiv191207990H}. From the algorithmic perspective, the authors in \cite{9104305} developed a channel estimation algorithm based on the parallel factor decomposition. The authors in \cite{8879620} introduced a general two-stage estimation algorithm by formulating the channel estimation problem as bilinear sparse matrix factorization and matrix completion. 
By exploiting the channel properties in RIS-assisted communication systems, a series of methods have been developed in \cite{9133156, 2019arXiv191203619C, 2019arXiv191207990H}. By exploiting the sparsity of the channels, the authors in \cite{9133156, 2019arXiv191203619C} formulated the channel estimation problem as a sparse signal recovery problem. By utilizing the property that the RIS-to-BS channel is quasi-static and the RIS-to-device channel is fast-varying, the authors in \cite{2019arXiv191207990H} proposed a two-timescale channel estimation framework. Nevertheless, most of the existing works on the channel estimation for RIS-assisted communication systems \cite{9104305, 2019arXiv191203619C, 2019arXiv191207990H, 9130088} required the allocation of orthogonal pilot sequences to all the devices. However, in massive connectivity, the length of pilot sequences is usually much smaller than the number of IoT devices. As a result, it is generally impossible to allocate orthogonal pilot sequences to all IoT devices, which yields unique challenges. Hence, these works cannot be directly applied to RIS-assisted massive connectivity systems. 

In this paper, we consider the uplink transmission in an IoT network, where a multi-antenna BS serves a large number of single-antenna IoT devices with the assistance of an RIS. We adopt the grant-free random access scheme to support massive connectivity, where the devices are sporadically active. To fully unleash the potential of the RIS, the BS is required to detect active devices and separately estimate the channels of the RIS-to-device link and the RIS-to-BS link with non-orthogonal pilots, which is referred as a \emph{RIS-related activity detection and channel estimation} problem. Our main contributions are summarized as follows.
\begin{itemize}
        \item We propose a realistic channel model for the RIS-assisted IoT network with massive connectivity. Taking into account the physical propagation structure of wireless channels, we model the channel from the RIS to the BS follows the  geometric distribution. Due to limited scattering between the RIS and the BS, the number of spatial paths between them is usually small. Therefore, the RIS-to-BS channel can be represented as a sparse channel matrix under the virtual angular domain. On the other hand, the IoT devices are sporadically active at any time instant, which results in the sparsity of the device transmission pattern. By simultaneously exploiting the sparsity of both sporadic transmission and the RIS-to-BS channel, we formulate the RIS-related activity detection and channel estimation problem as a sparse matrix factorization problem.

        \item As the channel matrices in RIS-assisted IoT networks are high dimensional due to the massive number of IoT devices and passive elements at the RIS, the computational complexity to infer the active devices and two separate channels may be prohibitive. To tackle such a high-dimensional inference problem, we develop a unified framework based on the Bayesian inference framework to jointly detect the active devices and estimate the two separate channels. We calculate the posterior mean estimators on a factor graph via the canonical sum-product message passing algorithm. To further reduce computational complexities, we approximate the messages based on the AMP framework, which includes central-limit-theorem (CLT) and Taylor-series arguments. Moreover, the sparsity of both device transmission and the RIS-to-BS channels is exploited to enhance the estimation accuracy.
        
        \item We conduct extensive numerical experiments to verify the effectiveness of our proposed algorithm. Specifically, our algorithm outperforms the three-stage algorithm proposed in \cite{9054415} in terms of both activity detection and channel estimation accuracy. Furthermore, for channel estimation, our proposed algorithm can achieve the similar performance of Genie-aided MMSE estimator, which assumes all the active devices are known in advance. Finally, our experiments also reveal that massive MIMO can significantly improve the estimation accuracy in terms of both activity detection and channel estimation for RIS-assisted uplink transmissions.
\end{itemize}


\subsection*{Organization and Notations}
The remainder of this paper is organized as follows. Section II introduces the system model and the channel models. Section III formulates the RIS-related activity detection and channel estimation problem and describes the proposed AMP-based algorithm. Section IV presents extensive numerical results of the proposed algorithm followed by the conclusions in Section V.

Throughout this paper, the complex number sets is denoted by $\mathbb{C}$. Scalars, vectors and matrices denote regular letters, bold small letters, and bold capital letters, respectively. The imaginary unit is denoted by $j \triangleq \sqrt{-1}$. We use superscripts $(\cdot)^*$, $(\cdot)\tran$ and $(\cdot)\herm$ to denote conjugate, transpose and conjugate transpose. The $(i,j)-$th entry of $\bm{X}$ is denoted by $x_{ij}$. $\bm{I}$ and $\diag{\bm{x}}$ denote the identity matrix and a diagonal matrix with diagonal entries specified by $\bm{x}$. $\mathbb{E}[\cdot]$ denotes the expectation operator and $\mathcal{CN}(\bm{x}; \bm{\mu},\bm{\Sigma})$ denotes the complex Gaussian distributions with mean $\bm{\mu}$ and variance $\sigma^2$.

\section{System Model}
\subsection{RIS-Assisted IoT Networks}\label{sec: system model}
\begin{figure}[htbp]
        \centering
        \includegraphics[scale=0.25]{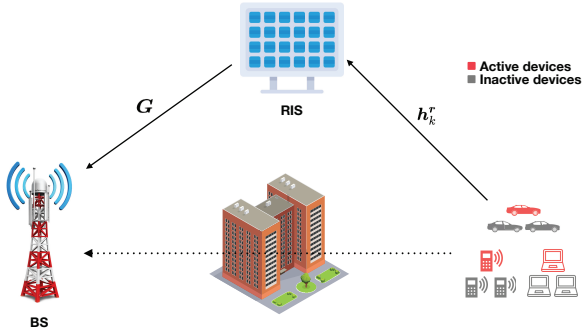}
        \caption{System model.}\label{fig: system model}
\end{figure}

In this paper, we consider the uplink transmission in an IoT network depicted in Fig. \ref{fig: system model}, where a BS equipped with $M$ antennas serves $K$ single-antenna IoT devices. An RIS consisting of $N$ passive reflecting elements is deployed to enhance the communication performance of IoT networks \cite{8811733}. Each reflecting element of the RIS is able to reflect the incident signals with desired phase shifts, which can be dynamically adjusted by the RIS controller \cite{8910627}. Due to the channel qualities of the direct links between the BS and the devices are much weak than that of the BS-RIS-device links. Hence, we follow \cite{8796365, 8741198, 9066923, 9133130}, and assume that the direct links between the BS and devices are not available.

In this paper, we adopt the grant-free random access scheme to support massive connectivity, where the transmission of IoT devices is sporadic \cite{8323218}. Each device is assumed to be active in each coherence block of length $T$ with probability $\lambda_\alpha$. Hence, only a subset of the devices are active within each transmission. In particular, the activity of the $k$-th device is indicated as follows
\begin{align}
        \alpha_k = \left\{ 
        \begin{array}{ll}
                1,\quad &\text{if device $k$ is active}, \\
                0,\quad &\text{otherwise},
        \end{array}
        \right.
        \forall k.
\end{align}
Hence, we have $\Pr(\alpha_k = 1) = \lambda_\alpha$. Furthermore, we define the support set of active devices as 
\begin{align}
        \mathcal{A} = \{k|\alpha_k = 1, 1\leq k \leq K\}.
\end{align}

For the purpose of detecting active devices and estimating the corresponding channels for the RIS-assisted IoT network, the $k$-th device is assigned to a unique signature sequence $\bm{q}_k \triangleq \left[q_{k,1}, \ldots, q_{k, L}\right]\tran \in \Cbb^{L\times 1}$, where the sequence length $L$ is typically smaller than coherence length $T$. We assume a block-fading channel model, where the channels are quasi-static in each coherence block. Namely, all the channels remain invariant in each coherence block, but vary independently over different blocks. We define $\bm{h}_k \in \Cbb^{N\times 1}$ and $\bm{G} \in \Cbb^{M\times N}$ as the $k$-th RIS-to-device channel and the RIS-to-BS channel, respectively.


In the $l$-th time slot, the received signal $\bm{y}_l \in \Cbb^{M\times 1}$ at the BS can be written as 
\begin{align}\label{system model1}
        \bm{y}_l = \sum_{k=1}^K\bm{G}\bm{\Phi}\bm{h}_k\alpha_k q_{k,l} + \nbf_l,
\end{align}
where $\nbf_l \sim \mathcal{CN}(\zerobf, \tau_{n}\bm{I})$ is the additive white Gaussian noise (AWGN) vector in the $l$-th time slot, and $\tau_{n}$ denotes the noise power. Besides, $\bm{\Phi} \triangleq \diag{\phi_1, \ldots, \phi_N} \in \Cbb^{N\times N}$ denotes the phase shift matrix of the RIS, where $\phi_n\in\Cbb$ denotes the phase shift of the $n$-th reflecting element. It is assumed that $|\phi_n| = 1, \forall n =1, \ldots, N$ and the phase of $\phi_n$ can be flexibly adjusted within $[0, 2\pi)$ \cite{8930608}.

Considering all $L$ time slots, the received signals $\Ybf = \left[\ybf_1,\ldots, \ybf_L\right]\in \Cbb^{M\times L}$ at the BS can be rewritten in the matrix form as
\begin{align}\label{system model2}
        \Ybf = \Gbf\Phibf\Hbf\Abf\Qbf + \Nbf,
\end{align}
where $\bm{H} \triangleq [\bm{h}_1, \ldots, \bm{h}_K]\tran \in \Cbb^{K\times N}$ is the RIS-to-device channel matrix, $\bm{Q} \triangleq [\bm{q}_1, \ldots, \bm{q}_K]\tran \in \Cbb^{K\times L}$ is the pilot matrix with $\qbf_k \triangleq [q_{k,1},\ldots, q_{k,L}]^{\tran}\in \Cbb^{L}$, $\Nbf \in \Cbb^{L\times M}$ is the independent AWGN and $\bm{A} \triangleq \diag{\alpha_1, \ldots, \alpha_K} \in \Cbb^{K\times K}$ is the activity matrix indicating the activity of devices.  In this paper, our goal is to detect the device activity $\alpha_k$ and estimate the corresponding channel vector $\bm{h}_k$ as well as the channel matrix $\bm{G}$, given the observations $\Ybf$, the known phase shift matrix $\bm{\Phi}$ and the known pilot matrix $\bm{Q}$ in the massive connectivity setting.

\subsection{Channel Models}\label{sec: channel models}
\subsubsection{RIS-to-BS Channels}
We assume the BS is equipped with a uniform linear array (ULA), while the RIS is equipped with an $N_1\times N_2$ uniform rectangular array (URA) with $N_1N_2 = N$. By exploiting the physical propagation structure of wireless channel, we consider the RIS-to-BS channel matrix $\Gbf$ as a geometric channel \cite{8316940}. By applying the geometric channel model \cite{tse2005fundamentals}, the RIS-to-BS channel can be expressed as
\begin{align}\label{eq: G}
        \Gbf=\tau_{G} \sqrt{\frac{MN}{P}}\sum_{p=1}^{P}  \kappa_p \abf_B\left(\theta_p\right) \abf_R\herm \left(\psi_p, \omega_p\right),
\end{align} 
where $P$ is the total number of spatial paths between the BS and the RIS, $\kappa_p$ is the complex-value channel gain of the $p$-th RIS-to-BS path; $\tau_{G}$ is the distance-dependent path loss, $\theta_p$ is the corresponding azimuth angle-of-arrival (AoA) at the BS, $\psi_p$ ($\omega_p$) denote the corresponding azimuth (elevation) angle-of-departure (AoD) at the RIS. We set $p=1$ to model the LoS path between the RIS and the BS. In addition, $\abf_{B}$ and $\abf_{R}$ are the steering vectors associated with the BS and the RIS antenna geometry, i.e.,
\begin{align}
        \abf_B(\theta) &= \frac{1}{\sqrt{M}}\left[1, e^{-j \frac{2 \pi}{\rho} d \sin (\theta)}, \cdots, e^{-j \frac{2 \pi}{\rho} d(M-1) \sin (\theta)}\right]\tran,\\
        \abf_R(\psi, \omega) &= \abf_{R,v}(\psi, \omega) \otimes \abf_{R,h}(\psi, \omega) ,
\end{align}
where $\rho$ is the carrier wavelength and $d$ is the antenna spacing. Here, $\mathbf{a}_{R, v}(\psi, \omega)$ and $\mathbf{a}_{R, h}(\psi, \omega)$ are the horizontal and vertical steering vectors defined as 
\begin{align}
        &\bm{a}_{R, h}(\psi, \omega) \nonumber \\
        &=\frac{1}{\sqrt{N_{1}}}\left[1, e^{-j \frac{2 \pi d}{\rho} f_h(\psi,\omega)}, \cdots, e^{-j \frac{2 \pi d}{\rho}\left(N_{1}-1\right) f_h(\psi,\omega)}\right]\tran,\\
        &\bm{a}_{R, v}(\psi, \omega) \nonumber \\
        &=\frac{1}{\sqrt{N_{2}}}\left[1, e^{j \frac{2 \pi d}{\rho} f_v(\psi,\omega)}, \cdots, e^{j \frac{2 \pi d}{\rho}\left(N_{2}-1\right) f_v(\psi,\omega)}\right]\tran,
\end{align}
where $f_h(\psi,\omega) = \cos (\omega) \sin (\psi)$ and $f_v(\psi,\omega) = \cos (\omega) \cos (\psi)$.

\subsubsection{RIS-to-Device Channels}
For the channels between the RIS and the devices, since the devices are generally surrounded by many reflective objects at low elevations in the dense urban environment, there are a rich number of local scatterers at the device side \cite{8316940}. On the other hand, the dense buildings and other objects are likely to block the LoS links between the RIS and the devices. Based on the above facts, we consider Rayleigh fading channels between the RIS and the devices, as in \cite{9066923, 9133130},
, i.e., each entry of $\hbf_k$ follows i.i.d complex Gaussian distribution:
\begin{align}\label{eq: h}
        p(h_{nk}) = \mathcal{CN}(h_{nk};0, \tau_{h,k}),
\end{align}
where $\tau_{h,k}$ is the distance-dependent path loss for the $k$-th RIS-to-device link .

\subsection{Problem Formulation} \label{sec: problem formulation}
Based on the observations $\Ybf$, the predetermined phase shift matrix $\bm{\Phi}$ and the known pilot matrix $\bm{Q}$, our goal is to detect the activity of devices $\{\alpha_k\}$ and estimate the corresponding channel vectors $\{\bm{h}_k\}$ as well as the channel matrix $\bm{G}$ in the massive connectivity regime. In this scenario, the sequence length is typically much smaller than the number of devices, i.e., $L \ll K$. Hence, it is impossible to assign the mutually orthogonal signature sequences to all devices. Inspired by the prior works \cite{8264818, 8536396}, we generate the signature sequence $\bm{q}_k$ according to i.i.d. complex Gaussian distribution with zero mean and variance $1/L$, i.e.,
\begin{align}\label{eq: sequence}
        \bm{q}_k \sim \mathcal{CN}\left(\zerobf, \frac{1}{L}\bm{I}\right).
\end{align}
Note that each signature sequence is normalized to have unit norm, i.e., $\mathbb{E}\left[\|\bm{q}_k\|_2\right] = 1, \forall k$.

All the reflecting elements of the RIS are switched on and set to have the same phase shift during the period of activity detection and channel estimation \cite{9133156}. By setting $\phi_n = 1, \forall n$, (\ref{system model2}) can be rewritten as
\begin{align}\label{system model3}
        \Ybf = \Gbf\Hbf\Abf\Qbf + \Nbf.
\end{align}
 
Recognizing that the RIS-to-BS channels are typically sparse due to limited scattering between the BS and the RIS, we represent the RIS-to-BS channel matrix $\Gbf$ on the virtual angular domain. In the following, by simultaneously exploiting the sparsity of both sporadic transmission and the RIS-to-BS channel, we show that the RIS-related channel estimation and activity detection problem can be formulated as a sparse matrix factorization problem.

\subsubsection{Sparsity of Sporadic Transmission}
By defining $\Xbf \triangleq \Hbf\Abf$, the received signals can be expressed as 
\begin{align}
        \Ybf = \Gbf\Xbf\Qbf + \Nbf.
\end{align}
Due to the sporadic transmission, only a few devices are active at the same time. Recall that the diagonal entry $\alpha_k$ of the matrix $\Abf$ indicates whether the $k$-th device is active or not, the activity matrix $\Abf$ is diagonal sparse, which further implies that the matrix $\Xbf$ is a column sparse RIS-to-device channel matrix. Note that the non-zero columns of $\Xbf$ represent the channel vector between the RIS and the active devices. Hence, we have
        \begin{align}\label{eq: xnk}
                x_{nk} = \left\{
                \begin{array}{ll}
                        h_{nk}, & \alpha_k = 1, \\
                        0, & \alpha_k = 0,
                \end{array}
                \right.
        \end{align}
where $x_{nk}$ is the $(n,k)$-th entry of matrix $\Xbf$.

\subsubsection{Sparsity of RIS-to-BS Channel Representation}
As both the RIS and the BS are typically installed at high elevations, there are only limited scattering between the BS and the RIS. According to \cite{2019arXiv191203619C, 9133156}, we represent the RIS-to-BS channel $\Gbf$  on the \emph{virtual angular domain} (VAD) to provide a discrete approximation of the physical channel. Instead of taking AoDs and AoAs from arbitrary physical angles, the VAD representation parameterizes them with pre-discretized angles with finite resolutions. Specifically, we employ three pre-discretized sampling grids $\bm{\vartheta}$ with length $M'$, $\bm{\varphi}$ with length $N'_1$ and $\bm{\varpi}$ with length $N'_2$ to parameterize $\{\theta_p\}_{1\leq p\leq P}$, $\{\psi_p\}_{1\leq p\leq P}$ and $\{\omega_p\}_{1\leq p\leq P}$, respectively. To well approximate the original channel \eqref{eq: G}, the angular resolutions should be large enough, i.e., $M' > M$, $N'_1 > N_1$ and $N'_2 > N_2$. Following by \cite{7227017}, the RIS-to-BS channel $\Gbf$ in \eqref{eq: G} can be expressed by 
\begin{align}\label{G under VAD}
        \Gbf = \Abf_{B}\Sbf (\Abf_R)\herm,
\end{align}
 where $\Abf_R = \Abf_{R,v}\otimes \Abf_{R,h}$. Here, we define $\Abf_B = \left[\abf_B(\vartheta_1), \ldots, \abf_B(\vartheta_{M'})\right] \in \Cbb^{M\times M'}$ is an over-complete matrix with each column representing a steering vector parameterized by the pre-discretized azimuth AoA at BS, and $\Abf_{R,h} = \left[\abf_{R,h}(\varphi_1), \ldots, \abf_{R,h}(\varphi_{N'_1})\right]\in \Cbb^{N_1\times N'_1}$ (or $\Abf_{R,v} = \left[\abf_{R,v}(\varpi_1), \ldots, \abf_{R,v}(\varpi_{N'_2})\right]\in \Cbb^{N_2\times N'_2}$) is an over-complete matrix with each column a steering vector parameterized by pre-discretized azimuth (or elevation) AoD at RIS, respectively. In addition, $\Sbf\in\Cbb^{M'\times N'}$ is the channel coefficient matrix in angular domain, where the non-zero $(i,j)$-th entry corresponds to the complex gain on the channel consisting of the $i$-th cascaded AoA array steering vector at the BS and the $j$-th AoD array steering vector at the RIS.

Since there are limited scatters between the BS and the RIS, the number of spatial paths between the BS and the RIS should be small, i.e., $P \ll \min\{M, N\}$. Therefore, only a few entries of $\Sbf$ are non-zero, i.e., $\Sbf$ is a sparse matrix. By substituting  \eqref{G under VAD} to \eqref{system model3}, the system model can be rewritten as 
\begin{align}\label{system model4}
        \Ybf = \Abf_B\Sbf\Abf_R^{\herm} \Xbf \Qbf + \Nbf.
\end{align}

With the distorted signals $\Ybf$, our goal is to recover the sparse coefficient matrix $\Sbf$ in angular domain and the sparse RIS-to-device channel matrix $\Xbf$, given the known pilot matrix $\Qbf$ and the predetermined matrices $\Abf_B$ and $\Abf_R$. Once we obtain the estimated matrix $\Xbfhat$, the diagonal entry of activity matrix $\Abfhat$ can be estimated via the group sparsity of $\Xbf$ as follows
\begin{align} \label{threshold}
    \hat{\alpha}_k = \left\{\begin{array}{l}
            1, \|\xbfhat_k \|_2 >  \epsilon\\
            0,  \|\xbfhat_k\|_2 \leq  \epsilon
    \end{array}\right.\quad 1\leq k \leq K,
\end{align}
where $\epsilon$ is a small positive threshold and $\xbfhat_k$ is the $k$-th column of the estimated matrix $\Xbfhat$ \cite{8536396}.

\subsection{Problem Analysis}

In terms of the activity detection and channel estimation problem in the massive connectivity scenario, the RIS-assisted system is quite different from the conventional communication systems, where the devices are usually served by the BS directly. In particular, in the conventional massive connectivity scenario, the BS only needs estimate the channels between the BS and the active devices. Due to the sporadic nature of IoT devices, channel estimation and activity detection can be achieved in a joint manner. For example, \cite{8536396} proposed a structured group sparsity estimation approach by exploiting sparsity in the device activity pattern. \cite{8323218} formulated the problem as the multiple measurement vector (MMV) problem and solved it via the AMP algorithm. In contrast, in the RIS-assisted scenario, since designing phase-shift matrix $\bm{\Theta}$ requires the knowledge of the RIS-to-BS channel matrix $\Gbf$ and the sparse RIS-to-device channel matrix $\Xbf$ separately \cite{8811733, 8930608, 9117136}, we need to decouple the RIS-related cascaded channels $\Gbf\Xbf$ given the pilot matrix $\Qbf$ and the observations $\Ybf$. This means that we cannot simply formulate the problem as the MMV problem. Moreover, without RF chains, the RIS cannot transmit pilot sequences and process received signals to help the BS detect activity and estimate channels \cite{8796365}. Hence, the BS will bear all the tasks of estimating and detecting.


On the other hand, for the RIS-assisted communication system, previous works \cite{9087848, 2019arXiv191203619C,  9133156} studied the channel estimation problem without considering the massive connectivity regime, all active devices are assumed to be known at the BS. By assuming that the number of active devices is smaller than the sequence length, i.e., $K < L$, they assign mutually orthogonal pilot sequences to all devices \cite{2019arXiv191203619C}. However, in the RIS-related massive connectivity regime, there are a large number of devices in the communication system, which raises unique computational challenges when solving the RIS-related channel estimation problem with limited time budget. In addition, it is infeasible to assign orthogonal pilot sequences to all devices. All these facts impose the critical challenges of channel estimation in the RIS-related massive connectivity scenario.

To tackle the above challenges brought by the RIS-related massive connectivity system, our previous work \cite{9054415} formulated the RIS-related activity detection and channel estimation problem as sparse matrix factorization, matrix completion and multiple measurement vector problem and proposed a three-stage algorithm. However, the proposed algorithm in \cite{9054415} requires the accuracy guarantee of each stage to ensure the convergence. To address this issue, this paper further proposes to formulate the RIS-related activity detection and channel estimation problem as a sparse matrix factorization problem given prior knowledge of the channels. Specifically, we concurrently leverage the channel sparsity of $\Gbf$ and the group sparsity of $\Xbf$. By utilizing the Bayesian inference framework, we propose a unified AMP based framework to efficiently jointly estimate the sparse channel coefficient matrix $\Sbf$ and the sparse RIS-to-device channel matrix $\Xbf$ in \eqref{system model4}. 

\section{AMP-Based RIS-related Activity Detection and Channel Estimation Algorithm}
In this section, we introduce an AMP-based algorithmic framework to solve the sparse signal recovery problem. We first reformulate the RIS-related activity detection and channel estimation problem as a Bayesian inference problem. To compute the optimal minimum mean-squared error (MMSE) estimate of $\Sbf$ and $\Xbf$, we derive the posterior probabilities of $\Sbf$ and $\Xbf$ conditioned on $\Ybf$ and represent those probabilities with a factor graph. Subsequently, we calculate these quantities by utilizing \emph{sum-product algorithm} (SPA) \cite{910572}. Furthermore, to implement the SPA in practice, we introduce some approximations to the SPA based on the central-limit-theorem (CLT) and Taylor-series arguments.

\subsection{Minimum Mean-squared Error (MMSE) Estimators}
In Bayesian inference framework, we treat $\Sbf$ and $\Xbf$ as random variables with known separable probability distribution functions (PDFs) $p(\Sbf)$ and $p(\Xbf)$. Hence, the Bayesian approach can exploit prior knowledge to further improve the estimation accuracy. According to \cite{Kay:2012069}, the optimal Bayesian estimator is the MMSE estimator. The MMSE estimators of $\Sbf$ and $\Xbf$ are given by $\hat{\Sbf} = \left[\hat{s}_{m'n'}\right]$ and $\hat{\Xbf} = \left[\hat{x}_{nk}\right]$, where
\begin{equation}\label{eq: MMSE estimators}
\begin{aligned}
        \hat{s}_{m'n'} &= \int s_{m'n'} p\left({s}_{m'n'}|\Ybf \right) \der s_{m'n'}, \\
        \hat{x}_{nk} &= \int x_{nk} p\left(x_{nk} | \Ybf \right) \der x_{nk}.
\end{aligned}
\end{equation}
The marginal posteriors with respect to $s_{m'n'}$ and $x_{nk}$ are derived as follows,
\begin{align}\label{eq: marginal posteriors}
        p\left({s}_{m'n'}|\Ybf \right) &= \int\int p\left(\Sbf, \Xbf | \Ybf\right)\der\Xbf \der \left(\Sbf \backslash s_{m'n'}\right) \text{and} \\
        p\left({x}_{nk}|\Ybf \right) &= \int\int p\left(\Sbf, \Xbf | \Ybf\right)\der\Sbf \der \left(\Xbf \backslash x_{nk}\right)
\end{align}
where $\Mbf\backslash m_{ij}$ denotes the collection of the entries of matrix $\Mbf$ excluding the $(i,j)$-th one. The MMSE for $\Sbf$ and $\Xbf$ according to \eqref{eq: MMSE estimators} are defined as
\begin{align}\label{eq: posterior mean estimator}
        \mathrm{MMSE}(\Sbf) &\triangleq \frac{1}{M'N'}\mathbb{E}\left[\|\hat{\Sbf} - \Sbf\|^2_F\right];\\
        \mathrm{MMSE}(\Xbf) &\triangleq \frac{1}{M'N'}\mathbb{E}\left[\|\hat{\Xbf} - \Xbf\|^2_F\right].
\end{align}

According to Bayes' rule, we arrive at 
\begin{align}
        p(\Sbf, \Xbf | \Ybf) &= \frac{1}{p(\Ybf)} p(\Ybf|\Sbf, \Xbf)p(\Sbf) p(\Xbf) \nonumber \\
        &\propto p(\Ybf|\Sbf, \Xbf) p(\Sbf) p(\Xbf),
\end{align}
where $p(\Ybf|\Sbf, \Xbf)$ is the likelihood function of $\Sbf$ and $\Xbf$. To facilitate the calculation of the likelihood function $p(\Ybf|\Sbf, \Xbf)$, we introduce the following two auxiliary variables $\Wbf = \Abf_B\Sbf\Abf_R^{\herm} \Xbf$ and $\Zbf =  \Wbf \Qbf$, and the system model \eqref{system model4} is rewritten as 
\begin{align}\label{system model5}
        \Ybf = \Wbf\Qbf + \Nbf = \Zbf + \Nbf.   
\end{align}
We assume that  the likelihood function of $\Zbf$ is element-wise separable. Since the noise $\Nbf$ is independent AWGN and $z_{ml} = \sum_{k=1}^K w_{mk}q_{kl}$, then it can be expressed by
\begin{align}
        p(\Ybf|\Zbf) &= \prod_{m=1}^M\prod_{l=1}^L p(y_{ml}|w_{mk}, \forall k)\nonumber \\
        & = \prod_{m=1}^M\prod_{l=1}^L \mathcal{CN}(y_{ml};\sum_{k=1}^Kw_{mk}q_{kl}, \tau_n).
\end{align}
Similarly, the PDF of the matrix $\Sbf$ is assumed to be element-wise separable. Due to the sparsity of the matrix $\Sbf$, we adopt Bernoulli-Gaussian distribution as the prior distribution for the matrix $\Sbf$, i.e., 
\begin{align}
        &p(\Sbf) = \prod_{m'=1}^{M'} \prod_{n'=1}^{N'} p(s_{m'n'}) \nonumber\\
        &= \prod_{m'=1}^{M'} \prod_{n'=1}^{N'}\left((1-\lambda_s)\delta(0) + \lambda_s\mathcal{CN}(s_{m'n'};0, \tau_s)\right), \label{prior: S}
\end{align}
where $\lambda_s$ is the Bernoulli parameter, $\tau_s$ is the variance of the nonzero entries of $\Sbf$ and $\delta(\cdot)$ is the Dirac delta function.

Recalling that $\Xbf = \Hbf\Abf$, the columns of $\Xbf$ share the same sparsity. By separating the activity matrix $\Abf$ and the RIS-to-device channel matrix $\Hbf$, we decompose each entry $x_{nk}$ of the matrix $\Xbf$ as $x_{nk} = \alpha_kh_{nk}$.
To model the sparsity of the matrix $\Xbf$, we model $\alpha_k$ as a Bernoulli random variable  with $\Pr(\alpha_k=1) = \lambda_\alpha$. We assume that the PDF of $\Xbf$ is assumed to be column-wise independent. Therefore, the prior of the $\Xbf$ can be written as 
\begin{align}
                p(\Xbf) &= \prod_{n=1}^N \prod_{k=1}^K p(x_{nk})        \nonumber\\ 
                &= \prod_{k=1}^K \prod_{n=1}^N\left((1-\alpha_k)(1-\lambda_\alpha)\delta(0) \right. \nonumber \\
                &\left.+ \alpha_k\lambda_\alpha\mathcal{CN}(h_{nk};0,\tau_{h,k})\right). \label{prior: X}
\end{align}

Recalling the relations among $\Ybf, \Zbf, \Wbf, \Sbf,$ and  $\Xbf$ in \eqref{system model5}, the posterior distribution the posterior distribution of $\Sbf$ and $\Xbf$ conditioned on the observations $\Ybf$ is given as,
\begin{align}
        &p(\Sbf, \Xbf | \Ybf)\propto \prod_{m=1}^M\prod_{l=1}^L p(y_{ml}|w_{mk}, \forall k) \nonumber \\
        &\times \prod_{m=1}^M\prod_{k=1}^K p(w_{mk}\big|\sum_{n=1}^N g_{mn}x_{nk}) \prod_{n=1}^N \prod_{k=1}^K p(x_{nk})\nonumber \\
        &\times  \prod_{m=1}^M \prod_{n=1}^N p(g_{mn}|s_{m'n'}, \forall m', n') \prod_{m'=1}^{M'} \prod_{n'=1}^{N'} p(s_{m'n'}) . \label{eq: posterior probability}
\end{align}

In practice, exact expectations of $\Sbf$ and $\Xbf$ are generally prohibitive due to the high-dimensional integrations involved in the marginalization (see Eq. \eqref{eq: marginal posteriors}). However, these quantities can be efficiently approximated using loopy belief propagation (LBP) \cite{frey1998revolution}.

\begin{figure}[htbp]
        \centering
        \includegraphics[scale=0.65]{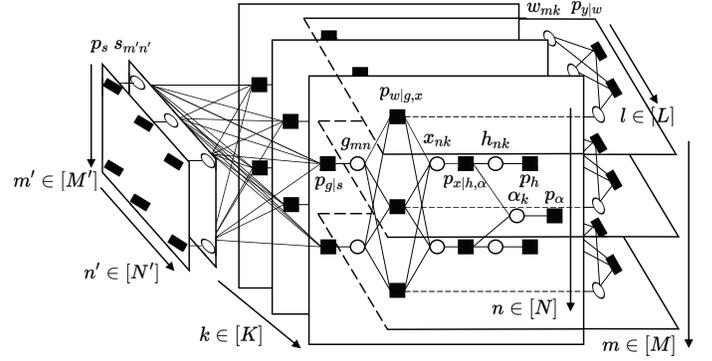}
        \caption{The factor graph for a toy-sized problem with dimensions $M'=N=L=2$ and $N'=M'=K=3$.}
        \label{fig: factor graph}
\end{figure}

\subsection{Loopy Belief Propagation}
\begin{table}[htbp]
        \begin{center}
        \caption{Notations for the factor graph in Fig. \ref{fig: factor graph}.} \label{table: notations for factor graph}
                \begin{tabular}{lll}
                \hline\hline
                Notation & Factor node & Distribution \\
                \hline
                $p_{y_{ml}|w_{mk}}$ & $p(y_{ml} | w_{mk}, \forall k)$ & $\mathcal{CN}(y_{ml};\sum_{k=1}^Kw_{mk}q_{kl}, \tau_n)$ \\
                $p_{w_{mk}|g_{mn},x_{nk}}$ & $p(w_{mk}|g_{mn}, x_{nk}, \forall n)$ & $\delta (w_{mk} - \sum_{n=1}^Ng_{mn}x_{nk})$      \\
                $p_{x_{nk}}$ & $p(x_{nk})$ & 
                \hspace{-2mm}$\begin{array}{l}
                (1-\alpha_k)(1-\lambda_\alpha)\delta(0) \\
                + \alpha_k\lambda_\alpha\mathcal{CN}(h_{nk};0,\tau_{h,k})
                \end{array}$ \\
                $p_{g_{mn}|s_{m'n'}}$ & $p(g_{mn} | s_{m'n'}, \forall m', n') $& 
                
                \hspace{-2mm}$\begin{array}{l}
                        \delta (g_{mn} - \sum_{m'=1}^{M'} \sum_{n'=1}^{N'} \\
                        a_{B,mm'}s_{m'n'}a_{R,nn'})
                \end{array}
                $ \\
                $p_{s_{m'n'}}$ & $p(s_{m'n'})$ & 
                \hspace{-2mm}$\begin{array}{l}
                        (1-\lambda_s)\delta(0) \\
                        +\lambda_s\mathcal{CN}(s_{m'n'};0, \tau_s)
                \end{array}$
                 \\
                \hline\hline
                \end{tabular}
        \end{center}
\end{table}

\begin{table}
        \begin{center}
        \caption{Notations of means and variances for the messages.}
        \label{table: notation for messages}
                \begin{tabular}{lll}
                        \hline\hline
                        Message & Mean & Variance \\
                        \hline
                        $\Delta_{ l\gets mk}^t(w_{mk})$ & $\what_{mk,l}(t)$& $v^w_{mk,l}(t)$\\
                        $\Delta_{k\gets mn}^t(g_{mn})$ & $\ghat_{mn,k}(t)$& $v^g_{mn,k}(t)$\\
                        $\Delta_{m \gets  nk}^t(x_{nk})$ & $\xhat_{nk,m}(t)$& $v^x_{nk,m}(t)$\\
                        $\Delta_{mn\gets  m'n'}^t(s_{m'n'})$ & $\shat_{m'n',mn}(t)$& $v^s_{m'n',mn}(t)$\\
                        $\Delta_{w_{mk}}^t(w_{mk})$ & $\what_{mk}(t)$& $v^w_{mk}(t)$\\
                        $\Delta_{g_{mn}}^t(g_{mn})$ & $\ghat_{mn}(t)$& $v^g_{mn}(t)$\\
                        $\Delta_{s_{m'n'}}^t(s_{m'n'})$ & $\shat_{m'n'}(t)$& $v^s_{m'n'}(t)$\\
                        $\Delta_{x_{nk}}^t(x_{nk})$ & $\xhat_{nk}(t)$& $v^x_{nk}(t)$\\
                         \hline\hline
                \end{tabular}
        \end{center}
\end{table}
We construct a factor graph shown in Fig. \ref{fig: factor graph} to represent the PDFs in \eqref{eq: posterior probability}. The random variables and the factors of the posterior probabilities in \eqref{eq: posterior probability} are represented by variable nodes appearing as while circles and factor nodes appearing as black squares, respectively. The notations of the factor nodes are summarized in Table \ref{table: notations for factor graph}. To simplify the notations, we omit the subscripts of the variable nodes in Fig. \ref{fig: factor graph}. By applying the traditional message passing algorithm, the estimators in \eqref{eq: posterior mean estimator} can be approximately computed. In high-dimensional inference problems, exact implementation of the SPA is impractical, which motivates approximations of the SPA. In the sequel, we will first derive the messages for the SPA, and then approximate these messages. As we shall see, the approximations are primarily based on central-limit-theorem (CLT) and Taylor-series arguments.

\subsection{Sum-Product Algorithm}
To employ the SPA, we define the following notations of messages: $\Delta^t_{ij\to mn}(x_{mn})$ denotes the message from the factor node $f_{ij}$ to the variable node $x_{mn}$ in the $t$-th iteration, $\Delta^t_{ij\to mn}(x_{mn})$ denotes the message from  the variable node $x_{mn}$ to the factor node $f_{ij}$, and $\Delta^t_{x}(x)$ denotes the marginal message computed at variable node $x$. 

By applying the SPA to the factor graph in Fig. \ref{fig: factor graph} and following the procedures in \cite{910572}, we obtain the following update rules of the messages:

\subsubsection{Messages between variable nodes $\{w_{mk}\}$ and factor nodes $\{p(y_{ml}|w_{mk}, \forall k)\}$ are given by}
\begin{align}
        &\Delta^{t}_{l\to mk}(w_{mk}) \propto \int p\left(y_{ml}|w_{mk},\forall k \right)       \prod_{j\neq k}\Delta^{t}_{l\gets mj}(w_{mj})\der w_{mj}, \label{message: f-w}\\
        &\Delta^{t+1}_{l\gets mk}(w_{mk}) \propto \mathcal{P}^t_{w_{mk}}(w_{mk})\prod_{j\neq l} \Delta^{t}_{j\to mk}(w_{mk}).
\end{align}
Here, the auxiliary distribution $\mathcal{P}^t_{w_{mk}}(w_{mk})$ denotes the messages from the variable nodes $\{g_{mn}\}$ and $\{x_{nk}\}$, which is defined as 
\begin{align}
        &\mathcal{P}^t_{w_{mk}}(w_{mk})         \propto \int p\left(w_{mk}|g_{mn}, x_{nk}, \forall n \right) \nonumber \\
        &\times \prod_{n=1}^{N}\Delta^t_{k\gets mn}(g_{mn}) \der g_{mn} \Delta^t_{m\gets nk}(x_{nk}) \der x_{nk} \label{message:pw}.
\end{align}

\subsubsection{Messages between variable nodes $\{g_{mn}\}$ and factor nodes $\{p(w_{mk}|g_{mn},x_{nk}, \forall n)\}$ are given by}
\begin{align}
        &\Delta_{k \to mn}^t(g_{mn}) \propto  \int \prod_{l=1}^L\Delta_{l \to mk}^t(w_{mk})p(y_{ml}|w_{mk},\forall k)\der y_{ml} \nonumber \\ 
        &\times \prod_{n=1}^{N}  \Delta_{m \gets nk}^t(x_{nk})\der x_{nk} \prod_{j \neq n} \Delta_{k\gets mj}^t(g_{mj})\der g_{mj}, \label{message: f-g}\\
        &\Delta_{k \gets mn}^{t+1}(g_{mn}) \propto \mathcal{P}^t_{g_{mn}}(g_{mn}) \prod_{j \neq  k}\Delta_{ j\to mn}^t(g_{mn}) \label{message: g-f}.
\end{align}
Here, the auxiliary distribution $\mathcal{P}^t_{g_{mn}}(g_{mn})$ denotes the messages from the variable nodes $\{s_{m'n'}\}$, which is defined as 
\begin{align}
&\mathcal{P}^t_{g_{mn}}(g_{mn}) \propto\int p(g_{mn} | s_{m'n'}, \forall m', n') \nonumber \\
&\times \prod_{m'=1}^{M'}\prod_{n'=1}^{N'} \Delta^t_{mn\gets m'n'}(s_{m'n'})\der s_{m'n'}. \label{message: pg}
\end{align}

\subsubsection{Messages between variable nodes $\{s_{m'n'}\}$ and factor nodes $\{p(g_{mn}|s_{m'n'}, \forall m', n')\}$ are given by}
\begin{align}
&\Delta_{mn\to m'n'}^t(s_{m'n'}) \propto \int p(g_{mn} | s_{m'n'}, \forall m', n')\nonumber\\
& \times \prod_{k=1}^{K} \Delta_{k \to mn}^t(g_{mn}) \der g_{mn} \prod_{(i,j)\neq (m',n')}\Delta_{mn \gets  ij}^{i}(s_{ij}) \der s_{ij}, \label{message: f-s}\\
&\Delta_{mn \gets  m'n'}^{t+1}(s_{m'n'}) \propto p(s_{m'n'})\prod_{(i,j)\neq (m,n)} \Delta_{ij \to m'n'}^t(s_{m'n'}) \label{message: s-f}.
\end{align}

\subsubsection{Messages between variable nodes $\{x_{nk}\}$ and factor nodes $\{p(w_{mk}|g_{mn}, x_{nk}, \forall n)\}$ are given by}
\begin{align}
        &\Delta_{m \to nk}^t(x_{nk}) \propto \int \prod_{l=1}^L\Delta_{l \to mk}^t(w_{mk})p(y_{ml}|w_{mk},\forall k)\der y_{ml} \nonumber \\
        &\times \prod_{n=1}^{N}\Delta_{k \to mn}^t(g_{mn})\der g_{mn} \prod_{j \neq n} \Delta_{m \gets jk}^t(x_{jk})\der x_{jk} , \label{message: f-x}\\
        &\Delta_{k \gets nk}^{t+1}(x_{nk}) \propto p(x_{nk}) \prod_{j \neq  m}\Delta_{j\to nk}^t(x_{nk}) \label{message: x-f}.
\end{align}

\subsubsection{Marginal messages at variable nodes are given by}
\begin{align}
\Delta_{w_{mk}}^{t+1}(w_{mk})&\propto \mathcal{P}^t_{w_{mk}}(w_{mk}) \prod_{l=1}^L\Delta^t_{l\to mk}(w_{mk}), \label{message: w}\\
\Delta_{g_{mn}}^{t+1}(g_{mn})&\propto \mathcal{P}^t_{g_{mn}}(g_{mn}) \prod_{k=1}^{K}\Delta_{k\to mn}^t(g_{mn}), \label{message: g}\\
\Delta_{s_{m'n'}}^{t+1}(s_{m'n'})&\propto p(s_{m'n'})\prod_{m=1}^{M}\prod_{n=1}^{N} \Delta_{mn \to m'n'}^t(s_{m'n'}). \label{message: s}\\
\Delta_{x_{nk}}^{t+1}(x_{nk})&\propto p(x_{nk}) \prod_{m=1}^M\Delta_{m \to nk}^t(x_{nk}). \label{message: x}
\end{align}

\subsection{Approximated Message Passing for SPA}
Due to the high-dimensional integrations, the messages in \eqref{message: f-w}--\eqref{message: s} are generally computationally intractable. Hence, we approximate the the SPA updates \eqref{message: f-w}--\eqref{message: s} based on the central-limit-theorem (CLT) and the Taylor-series arguments that are almost exact in the large-system limit, i.e., $M,M',N,N',K, L\to \infty$ with the fixed ratios $M/K$, $M'/K$, $N/K$, $N'/K$, and $L/K$, which is widely adopted in \cite{5695122, 6033942}. Specifically, we will neglect terms that vanish relative to others as $K\rightarrow \infty$, which is reasonable in massive connectivity. We first outline the main steps of the approximation method as follows, and the details of derivations can be referred to Appendix \ref{appendix: A}.

\begin{enumerate}[1.]
        \item We first adopt a second-order Taylor expansion to approximate $\mathcal{P}_{w_{mk}}^t$ as a Gaussian distribution. Due to the CLT arguments, we also approximate $\prod_{j\neq k}\Delta^t_{j \to mk}(w_{mk})$ as a Gaussian distribution. According to \eqref{message: f-w}, $\Delta^{t+1}_{l\gets mk}(w_{mk})$ is thus characterized as a Gaussian distributions with the tractable mean and variance. 
        
        \item We find that $\Delta_{l \gets mk}^t(w_{mk})$ differs from $\Delta_{w_{mk}}^t(w_{mk})$ in only one term $\Delta^t_{l\to mk}(w_{mk})$ that will vanish in the large-system limit. As a result, $\Delta_{w_{mk}}^t(w_{mk})$ also becomes a Gaussian distribution. Then we use the same mean $\what_{mk}(t+1)$ and variance $v^w_{mk}(t+1)$ to characterize these two messages. Consequently, the closed-loop updating formulas for $\what_{mk}$ and $v^w_{mk}$ can be obtained. As we shall see, the high-dimensional integration can be so that the computational complexity is significantly reduced.

        \item Similarly, we can show that $\Delta_{g_{mn}}^{t+1}(g_{mn})$, $\Delta_{x_{nk}}^{t+1}(x_{nk})$ and $\prod_{m,n}\Delta^t_{mn \to m'n'}(s_{m'n'})$ can be approximated as Gaussian distributions with corresponding means and variances as well. Taking the prior information \eqref{prior: S}--\eqref{prior: X} and the messages \eqref{message: s}--\eqref{message: x} into account, we can obtain the closed-loop updating formulas for $\ghat_{mn}(t)$, $v^g_{mn}(t)$, $\xhat_{nk}(t)$, $v^x_{nk}(t)$, and $\shat_{m'n'}(t)$, $v^s_{m'n'}(t)$. The notations of the messages are summarized in Table \ref{table: notation for messages}.
\end{enumerate}
The whole procedure is summarized in Algorithm \ref{algorithm}, and the algorithm will continue until meeting a convergence condition, i.e., the maximum number of iterations $I_{\text{max}}$ is reached.

\begin{algorithm}[t]
        \caption{The proposed algorithm.}
        \label{algorithm}
        \KwIn{$\Ybf; \Abf_B; \Abf_R; \Qbf; \tau_n; \tau_s; \tau_{h,k}; \lambda_s;\lambda_\alpha; \epsilon$}
        \KwOut{$\hat{\bm{S}}; \hat{\bm{X}}; \hat{\bm{A}}$}
        \Init{\\
        $\forall m', n'$: \emph{Sample} $\shat_{m'n'}(1)$ \emph{according to} $p(s_{m'n'})$\emph{;}\\
        $\forall n, k$: \emph{Sample} $\xhat_{nk}(1)$ \emph{according to} $p(x_{nk})$\emph{;} \\
        $\forall m, n$: $\ghat_{mn}(1) = \sum_{m'=1}^{M'}\sum_{n'=1}^{N'}a_{B,mm'}\shat_{m' n',mn}(1)a_{R,n' n}$\emph{;} \\
        \vspace{1mm}
        $\forall m, k$: $\what_{mk}(1) = \sum_{n=1}^N \ghat_{mn}(1) \xhat_{nk}(1)$\emph{;} \\
        }
        
        \For{$i=1,2,\cdots, I_{\mathrm{max}}$}{ 
        $\forall m, k$: update $\what_{mk}(t)$ and  $v^w_{mk}(t)$ via \eqref{update: w1}--\eqref{update: w2}, \eqref{update: w3}, and \eqref{update: p};\\
        $\forall m, n$: update $\ghat_{mn}(t)$ and  $v^g_{mn}(t)$ via \eqref{temp57}, \eqref{vb}--\eqref{bh}, and \eqref{update: x}; \\
        $\forall m', n'$: update $\shat_{m'n'}(t)$ and $v^s_{m'n'}(t)$ via \eqref{update: s1}-- \eqref{muh} and \eqref{update: g};\\
        $\forall n, k$: update $\xhat_{nk}(t)$ and  $v^x_{nk}(t)$ via \eqref{update: s1}--\eqref{dh} and \eqref{update: s};
        }
        Activity matrix $\hat{\bm{A}}=\diag{\hat{\alpha}_k}$ is determined by \\
        \begin{align*}
		    \hat{\alpha}_k = \left\{\begin{array}{l}
		            1, \|\xbfhat_k \|_2 >  \epsilon\\
		            0,  \|\xbfhat_k\|_2 \leq  \epsilon
		    \end{array}\right.\quad 1\leq k \leq K.
		\end{align*}
\end{algorithm}

\subsubsection*{Remark 1} The computational complexity of each step is shown in Table \ref{table: complexity}. The complexity of the proposed algorithm in each iteration is $\mathcal{O}(MK)+\mathcal{O}(MN)+\mathcal{O}(NK)+\mathcal{O}(MNM'N')$. Note that, with the fixed ratios $M/K$, $M'/K$, $N/K$, $N'/K$, and $L/K$, the overall complexity of the proposed algorithm can be simplified as $\mathcal{O}(IK^4)$. This polynomial scaling of the complexity with respect to problem dimensions gives encouragement that our algorithm should efficiently handle large-scale problems. In contrast, the exact MMSE estimators of $\Xbf$ and $\Sbf$ in \eqref{eq: MMSE estimators} involving integrations with respect to $\Xbf$ and $\Gbf$, therefore the corresponding computational complexity grows exponentially with $K^2$.

\subsubsection*{Remark 2} Although this paper models the channels as in Section \ref{sec: channel models}, the proposed algorithm still needs the exact parameters that determine these distributions, i.e., $\tau_n$. In general, these parameters are usually unobtainable. To tackle this problem, we can parameterize the priors in the proposed algorithm and exploit the expectation-maximization (EM) based approach to tune the parameter. Due to the space limitation, we will not go into the details of the method in this paper, which can be referred to \cite{6898015}.

\begin{table}
        \begin{center}
                \caption{Computational complexities of each step in Algorithm \ref{algorithm}.}\label{table: complexity}
                \begin{tabular}{ll}
                        \hline\hline
                        Step & Complexity \\
                        \hline\hline
                        Updating $\what_{mk}(t)$ and $v^w_{mk}(t)$ & $\mathcal{O}(MK)$\\
                        Updating $\ghat_{mn}(t)$ and $v^g_{mn}(t)$ & $\mathcal{O}(MN)$\\
                        Updating $\xhat_{nk}(t)$ and $v^x_{nk}(t)$ & $\mathcal{O}(NK)$\\
                        Updating $\shat_{m'n'}(t)$ and $v^s_{m'n'}(t)$ & $\mathcal{O}(MNM'N')$\\
                         \hline\hline
                \end{tabular}
        \end{center}
\end{table}

%

%
%

\section{Numerical Experiments}
\subsection{Simulation Setting}
\begin{figure}
        \centering
        \includegraphics[scale=0.45]{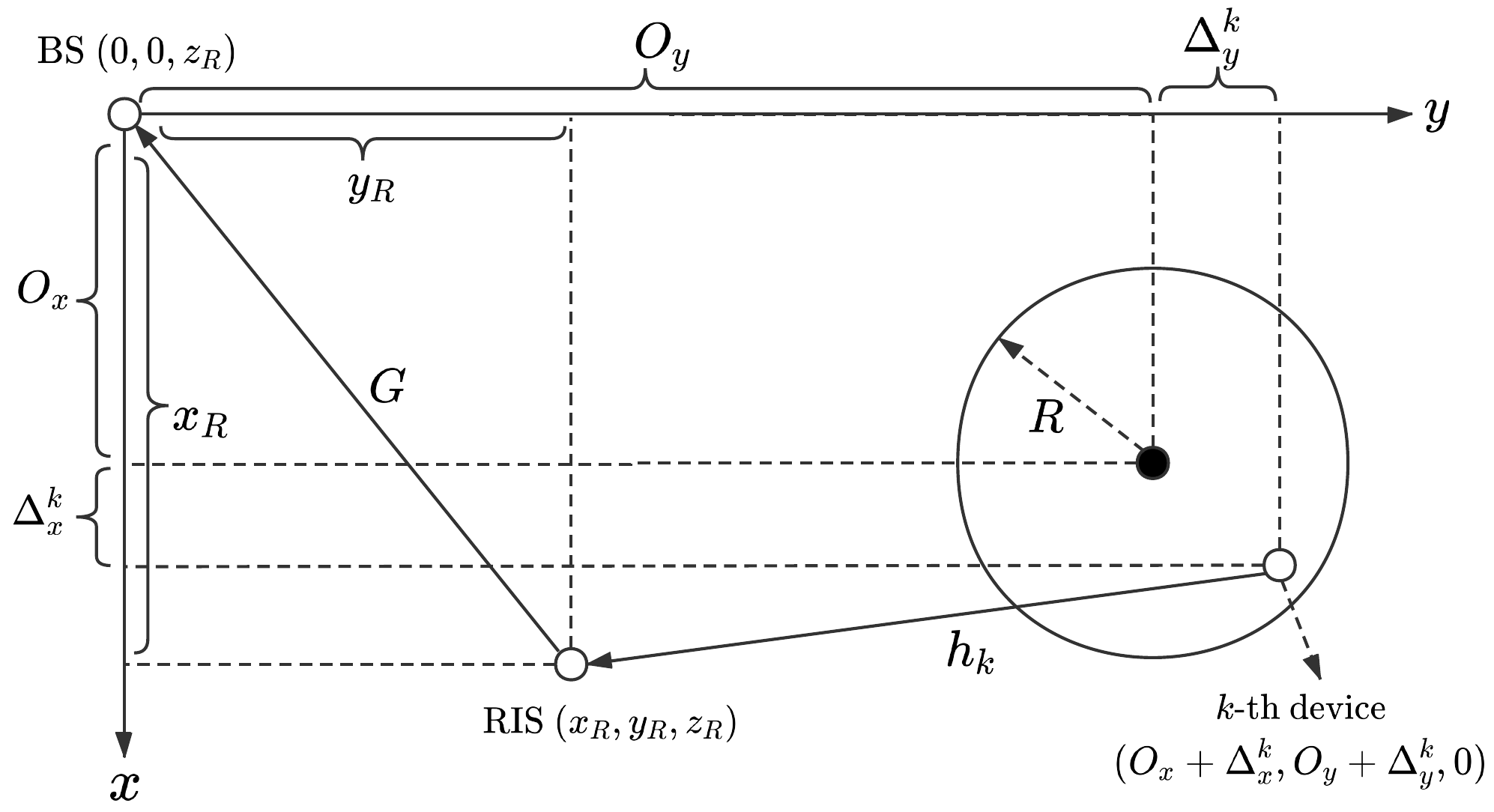}
        \caption{Horizontal locations of the BS, RIS and devices.}
        \label{fig: horizontal location}
\end{figure}
In this section, we conduct several numerical experiments to verify the effectiveness of the proposed algorithm. There are totally $K = 1000$ devices, where 80 of them are active, i.e., $\lambda_a = 0.08$. Under a three dimensional (3D) Cartesian coordinate system, we consider that the BS is equipped with a uniform linear array (ULA) aligned with $z$-axis; while the RIS is equipped with a uniform rectangular array (URA) parallel to the $x-z$ plane. For illustration, we assume that both the BS and the RIS are located at the same altitude above the devices by $z_R$ m. In addition, we assume that all devices are randomly and uniformly located in a circular coverage area of radius $R=50$ meters (m) with a center whose location is $(O_x, O_y, 0)$. The locations of the BS, the RIS and the $k$-th device are set as $(0,0,z_R)$, $(x_R, y_R, z_R)$ and $(O_x + \Delta_x^k, O_y + \Delta_y^k, 0)$, respectively, whose horizontal projections are illustrated in Fig. \ref{fig: horizontal location}. The 3D distances for the RIS-to-BS link and the $k$-th RIS-to-device link can be obtained as $d_G=\sqrt{x_R^2+y_R^2}$ m and $d_k=\sqrt{(O_x + \Delta_x^k)^2 + (O_y+\Delta_y^k)^2+z_R^2}$ m, respectively. For all numerical experiments, we set $z_R = 10$ m, and the horizontal distances from the projection of the RIS on the $y$-axis and $x$-axis to the RIS as $x_R = 5$ m and $y_R = 100$ m.

We consider the distance-dependent path loss for all channel, which is modeled as
\begin{align}
        \tau = \tau_0 \left(\frac{d}{d_0} \right)^{-\mu},
\end{align}
where $\tau_0 = -30$ dB denotes the path loss at the reference distance $d_0 = 1$ m followed by \cite{9110912}; $\mu$ denotes the path loss exponent. In our setting, $\tau_G$ and $\tau_{h,k}$ are denoted as the path loss of the RIS-to-BS link and the $k$-th RIS-to-device link, respectively; the path loss exponents for the corresponding links are set as $\mu_{G}=2.2$ and $\mu_{h,k}=2.5$, respectively. According to \eqref{eq: h}, the RIS-to-device channel matrix $\Hbf$ is modeled by Rayleigh fading with $h_{nk} \sim \mathcal{CN}(0, \tau_{h,k}), \forall k,n$. On the other hand, we generate the RIS-to-BS channel matrix $\Gbf$ by \eqref{eq: G} with 10 clusters of paths and 5 subpaths per cluster. We draw the central azimuth AoA at the BS of each cluster uniformly over $[-\pi/2, \pi/2]$; draw the central azimuth (or elevation) AoD at the RIS of each cluster uniformly over $[-\pi, \pi]$ (or $[-\pi/2, \pi/2]$); and draw each subpath with a $\pi/12$ angular spread. Moreover, every complex-value channel gain $\kappa_p$ is drawn from $\mathcal{CN}(0,1)$. In our experiments, we set the pre-discretized sampling grids $\mathbf{\vartheta, \varphi}$ and $\mathbf{\varpi}$ to be uniform sampling grids over their corresponding domain, and the length of the sampling grids are set to have a fixed ratio to the antenna dimensions, i.e., $M'/M = N'_1/N_1 = N'_2/N_2 = 2$. Unless specifically mentioned, we set other system parameters as $\lambda_a = 0.08, M = 40, N_1 = N_2 = 7 (N = 49)$ and $I_{\text{max}} = 2000$ for the proposed algorithm. All the numerical results are averaged over 50 independent channel realizations.

In the following, we provide numerical experiments to evaluate the performance of the proposed  algorithm from the aspects of activity detection and channel estimation.

\subsection{Simulation for Activity Detection}
\subsubsection{Performance Metric} We evaluate the algorithms with the following metric. We define the probabilities of false alarm and missed detection to evaluate the performance of the proposed algorithm for activity detection. In particular, the probability of false alarm, $p_F$ is defined as the probability that a device is active but the detector declares the device is inactive, and the probability of missed detection, $p_M$ is defined as the probability that a device is inactive but the detector declares the device is active.

\subsubsection{Baselines}
In addition to the proposed algorithm, we introduce the following two algorithms as baselines for comparison.

\begin{itemize}
        \item \emph{AMP-MMV:} Instead of estimating the matrices $\Gbf$ and $\Xbf$, we estimate the cascaded channel matrix $\Wbf = \Gbf\Xbf$, and then detect the activity matrix $\Abf$. This problem is formulated as a \emph{Multiple Measurement Vector} (MMV) problem which can be efficiently solved by the AMP-MMV algorithm \cite{6320709}. Since we do not estimate the matrices $\Gbf$ and $\Xbf$ separately, the solution of AMP-MMV should be the performance lower bound for the probabilities of false alarm and missed detection.
        \item \emph{Three-stage \cite{9054415}:} By designing the phase shift matrix $\bm{\Phi}$ as a sparse matrix, the problem can be formulated as \emph{sparse matrix factorization}, \emph{matrix completion} and \emph{MMV} three stages. The solution of activity detection can be achieved in the third stage.
\end{itemize}


\begin{figure}
\centering
\begin{subfigure}[b]{0.24\textwidth}
        \centering
        \includegraphics[scale=0.37]{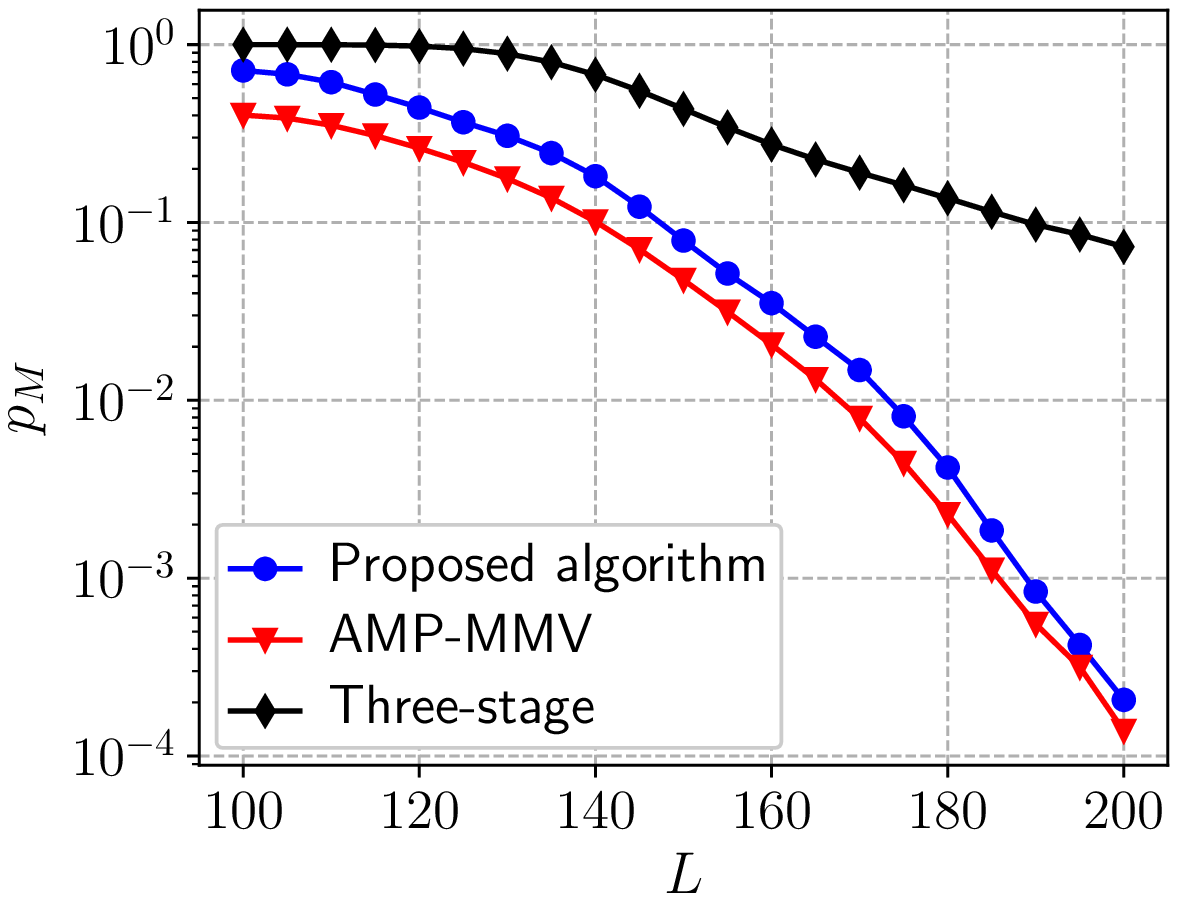}
        \caption{}
        \label{fig: prob_M}
\end{subfigure}
\begin{subfigure}[b]{0.24\textwidth}
        \centering
        \includegraphics[scale=0.37]{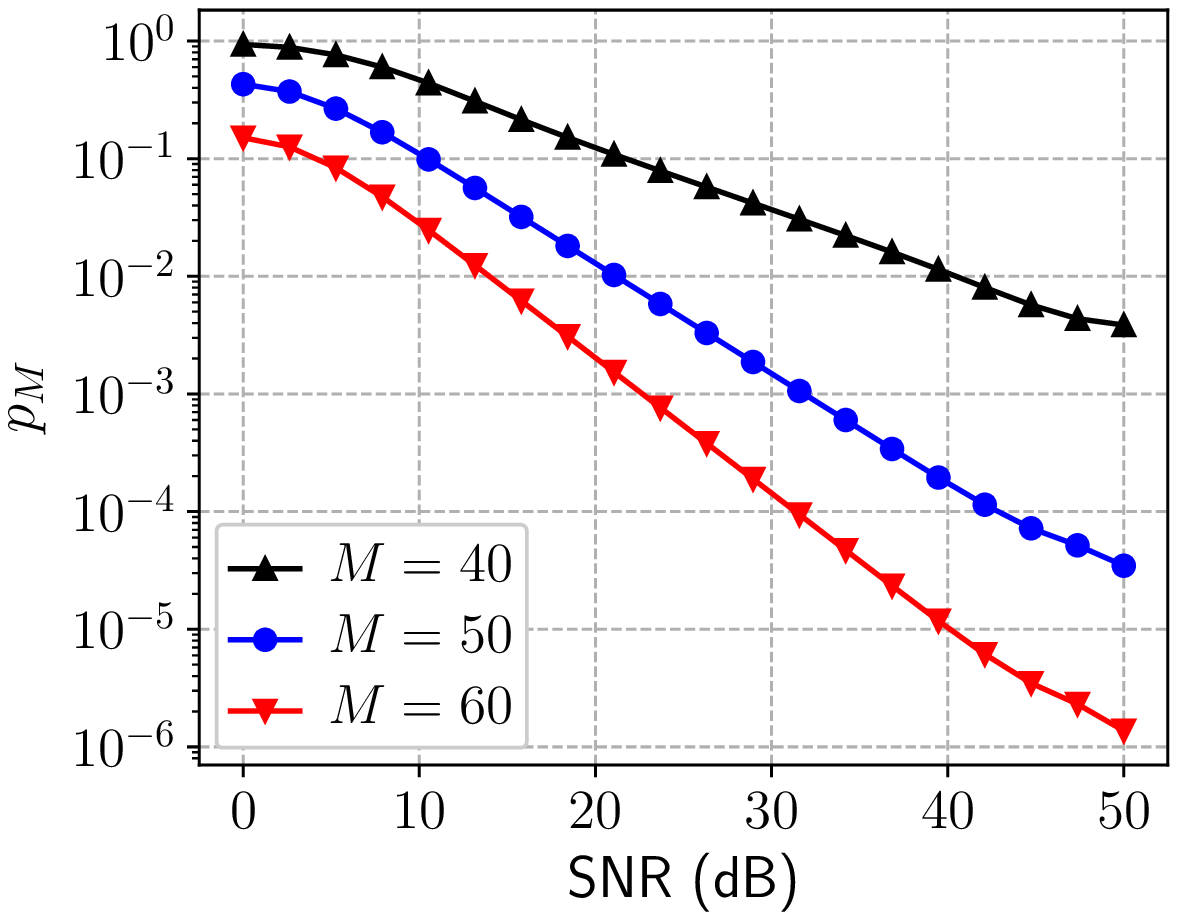}
        \caption{}
        \label{fig: prob2}
\end{subfigure}
\caption{The performance of $p_M$ with the fixed $p_F = 0.1$.}
\label{fig: prob}
\end{figure}

\subsubsection{Simulation Results}
Fig. \ref{fig: prob_M} investigates the impacts of sequence length $L$ on $p_M$ with the fixed $p_F = 0.1$. It can be shown that $p_M$ obtained by the proposed algorithm are very close to the ones obtained by the AMP-MMV algorithm, while the proposed algorithm estimates separately the matrix $\Gbf$ and $\Xbf$. Moreover, the proposed algorithm outperforms the three-stage algorithm \cite{9054415}, and the gap between the two algorithms increases when $L$ increases. Until the sequence length $L \geq 130$, the probabilities of the three-stage algorithm begin to decrease. The reason is that the estimated error in the first two stages will greatly affect the performance of activity detection. When the estimated error in the first two stages is relatively large, the third stage cannot detect activity accurately. In contrast, our algorithm provides a unifying framework instead of three-stage framework, and then is able to iteratively update all factor nodes in each message passing iteration, which avoids the problem arising in the three-stage algorithm. Fig. \ref{fig: prob2} shows the performance of activity detection as SNR varies for different number of antennas $M$ at the BS. By comparing the cases when $M=40$, $M=50$ and $M=60$ with the fixed sequence length $L = 130$ and the fixed $p_F = 0.1$, we observe that under the our proposed framework, $p_M$ decreases significantly as the number of BS antennas increases.

\subsection{Simulation for Channel Estimation}
\subsubsection{Performance Metric}
The performance of the proposed algorithm for channel estimation is evaluated by the normalized MSEs (NMSEs) of the RIS-to-BS channels $\Gbf$ and the average NMSEs of the active RIS-to-device channels $\{\hbf_k\}$, which are defined by
\begin{align}
        \text{NMSE of~}{\Gbf} &\triangleq \frac{\|\Gbf-\hat{\Gbf}\|^2_F}{\|\Gbf\|^2_F}, \\
        \text{Average NMSE of~}{\{\hbf_k\}} &\triangleq \frac{1}{|\mathcal{A}|} \sum_{k\in \mathcal{A}} \frac{\|\hbf_{k} - \hat{\hbf}_{k}\|_2^2}{\|\hbf_{k}\|_2^2}.
\end{align}
\subsubsection{Baselines}
We also consider two baseline algorithms for comparison. The three-stage algorithm \cite{9054415} has been introduced in the previous subsection. Here, we introduce Genie-aided MMSE estimator where all the active devices are assumed to be known at the BS in advance. Hence, the performance of Genie-aided MMSE estimator is referred as the bound of the NMSE performance for channel estimation. 

\subsubsection{Simulation Results}
\begin{figure}
 \centering     
 \begin{subfigure}[b]{0.24\textwidth}
        \centering
        \includegraphics[scale=0.37]{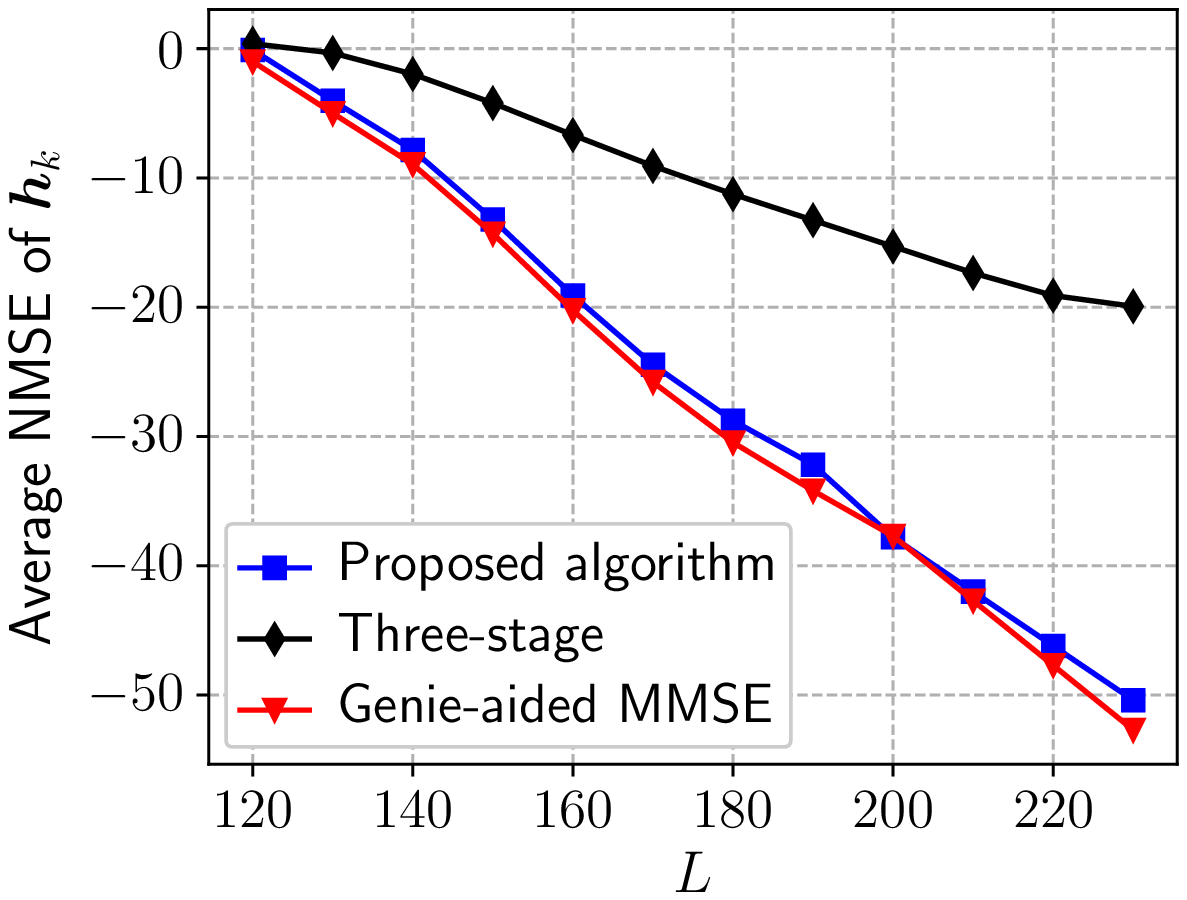}
        \caption{}
        \label{fig: ce1}
\end{subfigure}
\begin{subfigure}[b]{0.24\textwidth}
        \centering
        \includegraphics[scale=0.37]{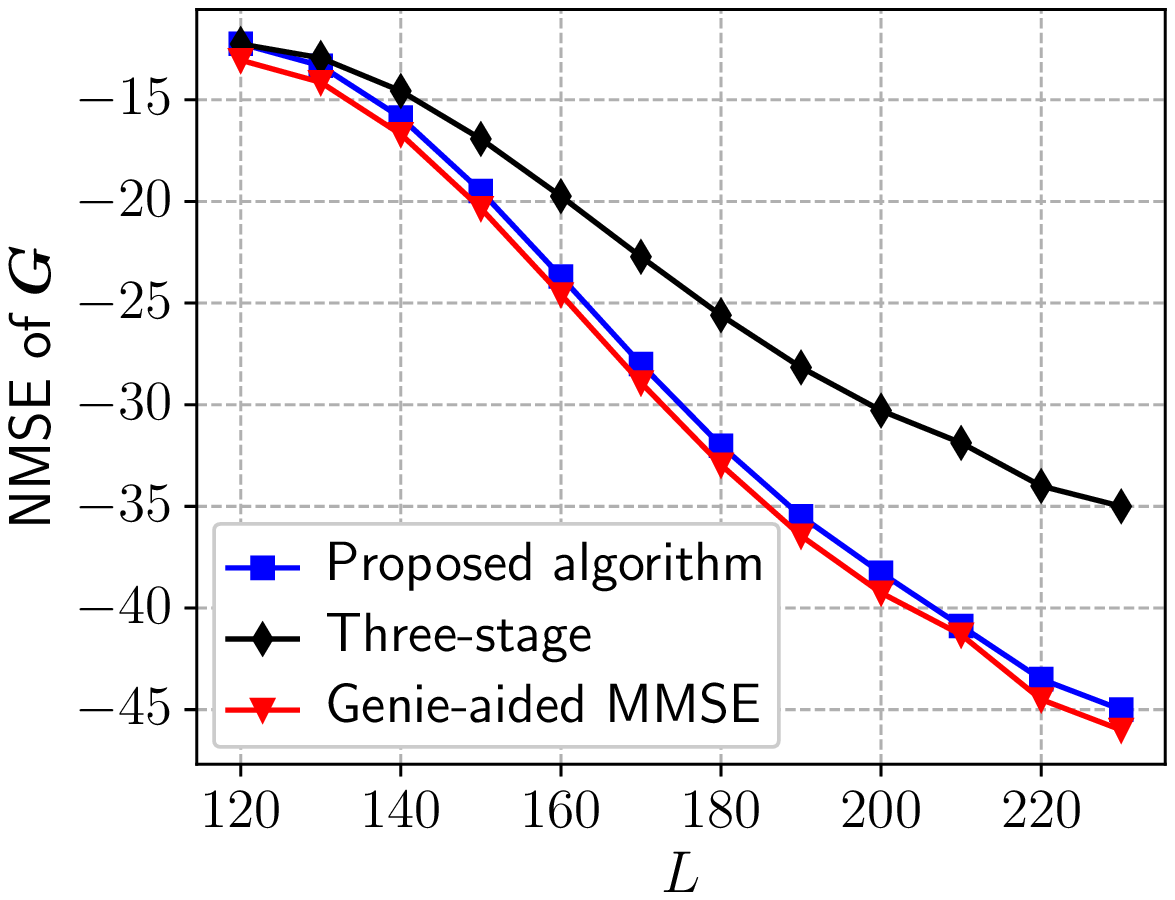}
        \caption{}
        \label{fig: ce2}
\end{subfigure}

\begin{subfigure}[b]{0.24\textwidth}
        \centering
        \includegraphics[scale=0.37]{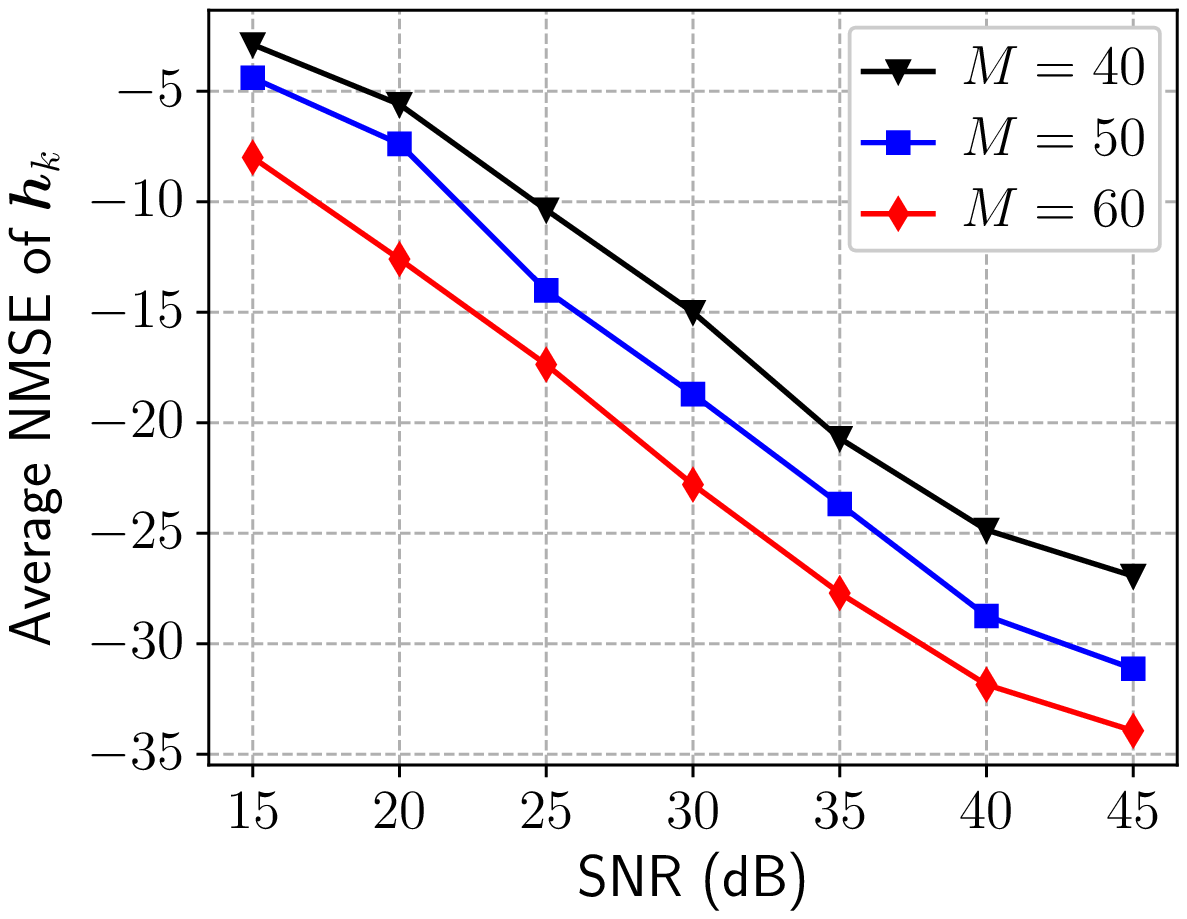}
        \caption{}
        \label{fig: ce3}
\end{subfigure}
\begin{subfigure}[b]{0.24\textwidth}
        \centering
        \includegraphics[scale=0.37]{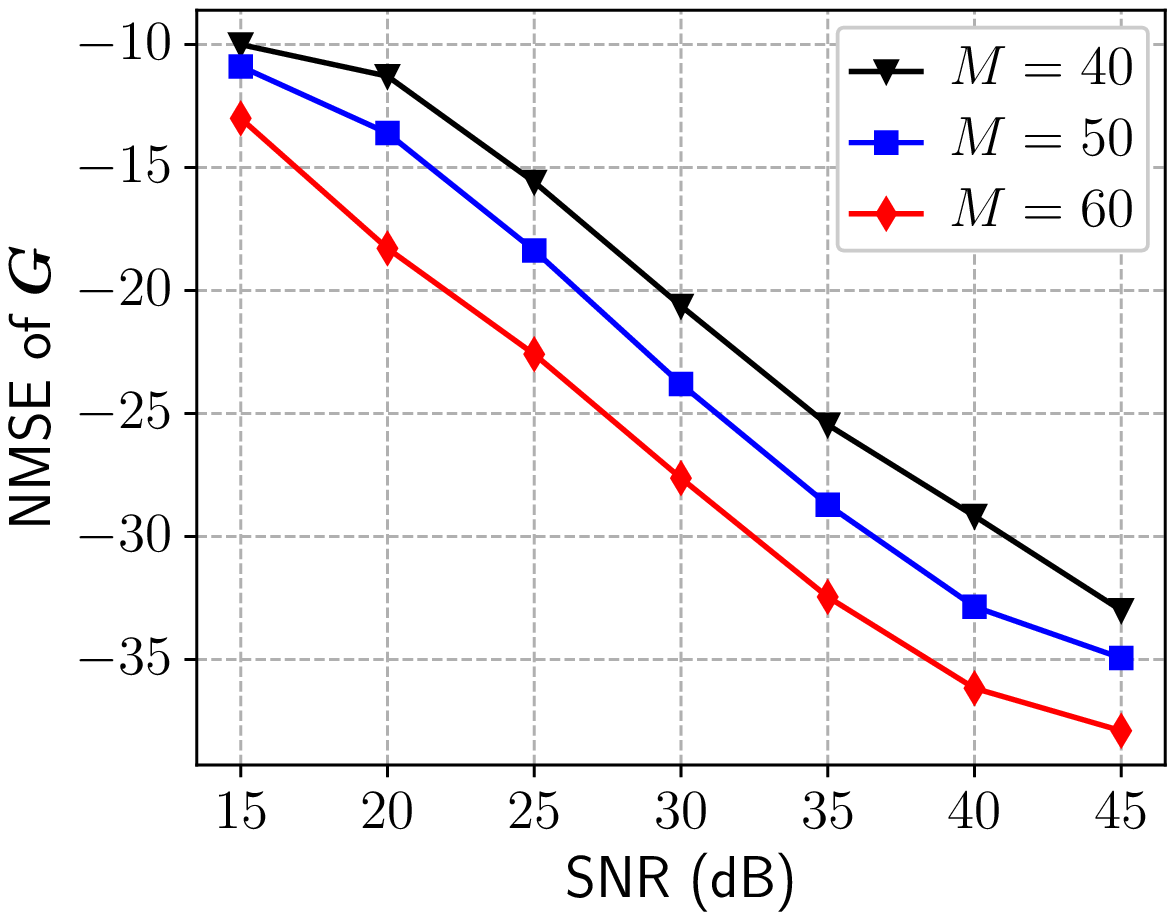}
        \caption{}
        \label{fig: ce4}
\end{subfigure}
\caption{The performance of channel estimation.}
\vspace{-5mm}
 \end{figure}

Fig. \ref{fig: ce1} and Fig. \ref{fig: ce2} show the effects of the sequence length on the NMSE of $\Gbf$ and the average NMSE of $\{\hbf_k\}$, respectively. We observe that the performance of the proposed algorithm can reach the bound (Genie-aided MMSE estimator) in both figures, namely, the proposed algorithm can correctly identified the active devices. In addition, we can see that the proposed algorithm achieves much lower NMSE of both $\Gbf$ and $\{\hbf_k\}$ than the three-stage algorithm over all sequence lengths, since the three-stage algorithm is limited to the correlation issue.  The main advantage of the proposed algorithm is to simultaneously exploit the sparsity of the RIS-to-BS channel $\Gbf$ and the sporadic transmission in massive connectivity.

On the other hand, we also investigate the channel estimation performance as SNR varies the cases when $M=40$, $M=50$ and $M=60$ with the fixed sequence length $L = 130$. We find that the NMSEs of the proposed algorithm decrease as SNR increases overall. Specifically, Fig \ref{fig: ce3} and Fig. \ref{fig: ce4} show that increasing the number of BS antennas $M$ brings significant improvement, i.e., for $M = 60$, the proposed algorithm approaches best performance in estimating $\Gbf$ and $\Hbf$ compared to other two cases. In other words, deploying massive antennas at the BS can significantly improve the estimation performance.

\begin{figure}
\centering
\begin{subfigure}[b]{0.24\textwidth}
        \centering
        \includegraphics[scale=0.32]{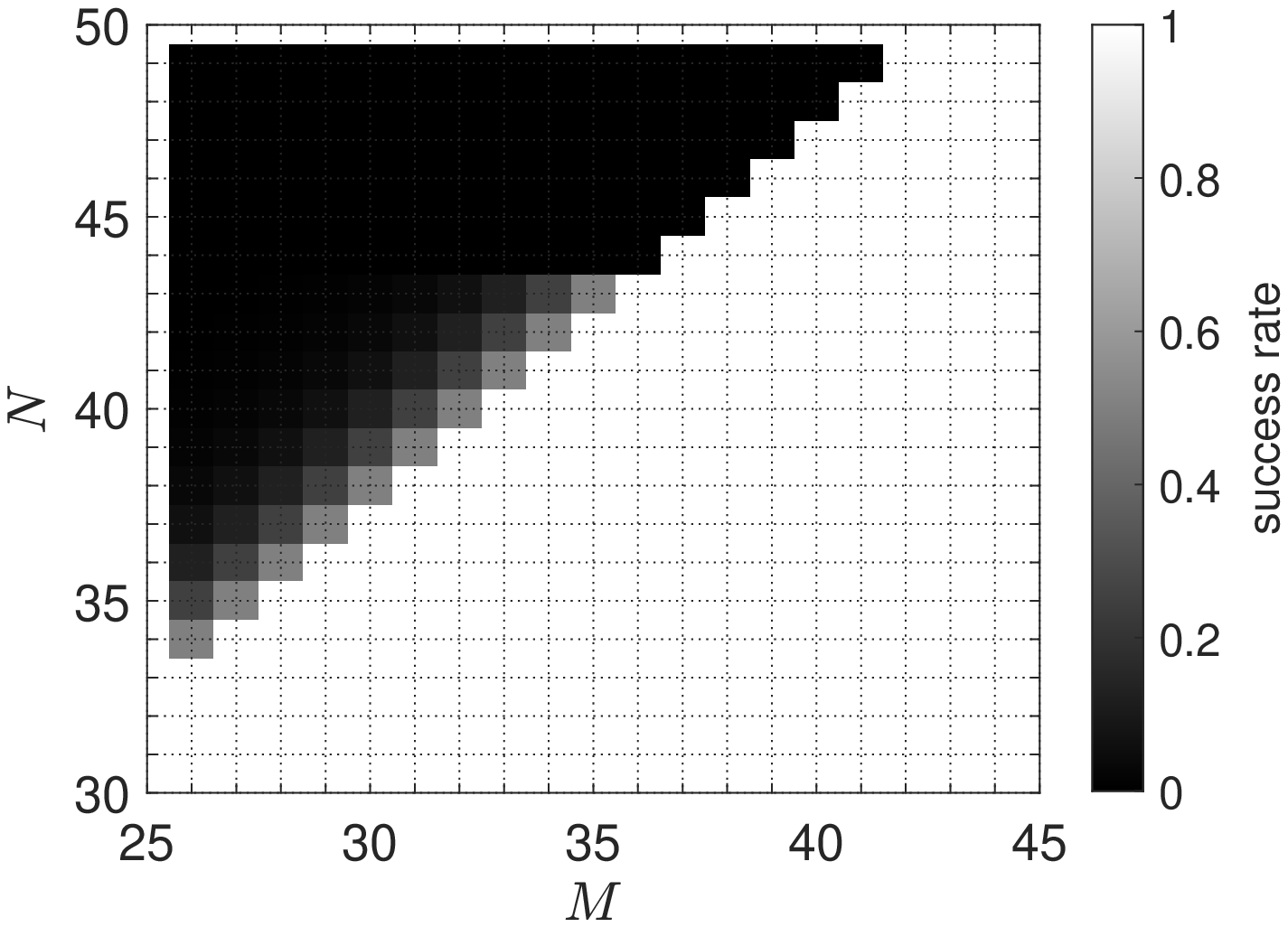}
        \caption{}
        \label{fig: pt1}
\end{subfigure}
\begin{subfigure}[b]{0.24\textwidth}
        \centering
        \includegraphics[scale=0.32]{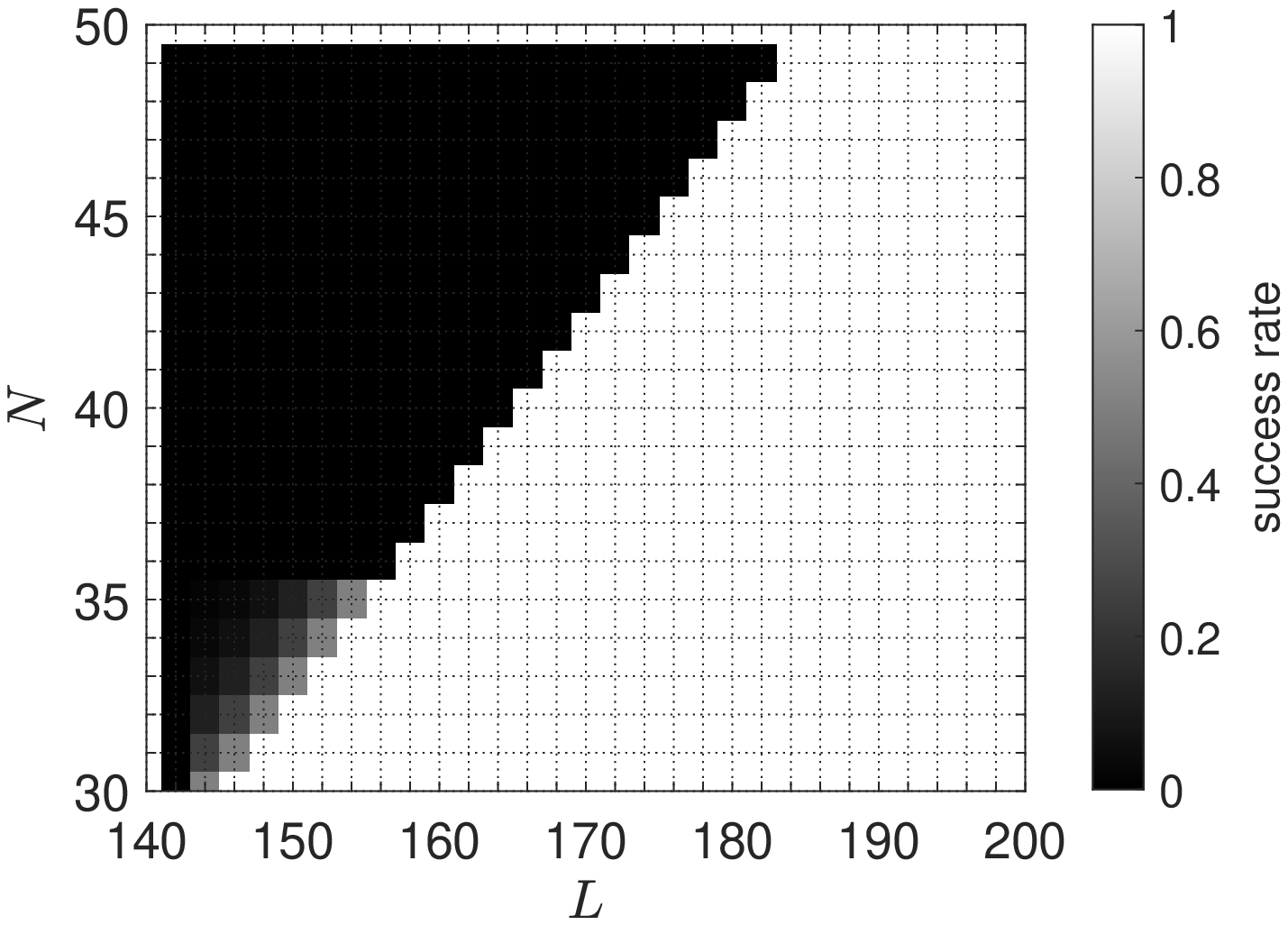}
        \caption{}
        \label{fig: pt2}
\end{subfigure}
\caption{The phase transitions of channel estimation.}
\label{fig: pt}
\vspace{-5mm}
\end{figure}

Finally, we demonstrate the performance limits of the proposed algorithm in terms of some critical systems parameters (i.e., $M$, $N$ and $L$). We declare the successful recovery of $\bm{G}$ and $\{\bm{h}_k\}$ if their NMSE are both less than $-30$dB. We perform 30 trials for each system setting to average and draw the phase transition diagrams in term of the success rate in Fig. \ref{fig: pt}. It illustrates that two sharp phase-transition curves separate success and failure regions. Based on the phase-transition curves, we can set appropriate system parameters for the proposed algorithm to efficiently support massive connectivity in RIS-assisted systems. For example, the minimal sequence length can be determined to accurately detect the active devices and estimate the channels.
 
\section{Conclusion}
In this paper, we studied the RIS-related activity detection and channel estimation problem in the RIS-assisted massive IoT network. We first modeled the RIS-to-BS channel as a sparse channel due to limited scattering between the RIS and the BS.  Consequently, we formulated the problem as a sparse matrix factorization problem by simultaneously exploiting the sparsity of both sporadic transmission in massive connectivity and the RIS-to-BS channels. Based on Bayesian inference framework, we further proposed an AMP-based algorithm to jointly estimate the detect active devices and two separated channels of the RIS-to-BS link and the RIS-to-device link. The computational complexity is further reduced by utilizing the central limit theorem and Taylor series arguments. Furthermore, we also conducted several numerical experiments to confirm the effectiveness and improvements of the proposed algorithm for the RIS-related activity detection and channel estimation problem.

\appendix[Details of Derivations] \label{appendix: A}
\subsection{Approximated messages of \eqref{message: f-w}-\eqref{message:pw}} Applying the Fourier inversion theorem and a second-order Taylor expansion to \eqref{message:pw}, we can approximate that $\mathcal{P}^t_{w_{mk}}(w_{mk}) \approx \mathcal{CN}(w_{mk}; \phat_{mk}(t), v^p_{mk}(t))$, where
\begin{subequations}\label{p}
        \begin{align}
                &v^p_{mk}(t) =\sum_{n=1}^{N}\left(\abs{\ghat_{mn,k}(t)}^2v^x_{nk,m}(t) \right. \nonumber \\
                & \left. +v^g_{mn,k}(t) \abs{\xhat_{nk,m}(t)}^2+v^g_{mn,k}(t)v^x_{nk,m}(t)\right), \\
                &\phat_{mk}(t)=\sum_{n=1}^{N} \ghat_{mn,k}(t)\xhat_{nk,m}(t). 
        \end{align}
\end{subequations}
The details can be referred to \cite[Eqs. (51)-(53)]{7457269}.

To approximate the message $\Delta^t_{l \to mk}(w_{mk})$, we define $z_{ml,k} \triangleq \sum_{j\neq k} w_{mj}q{jl} \sim \prod_{j\neq k}\Delta^t_{l\gets mj}(w_{mj})$. According to the CLT, we treat $z_{ml,k}$ as a Gaussian random variable with mean $\zhat_{ml}(t)-\what_{mk,l}(t)q_{kl}$ and variance $v^z_{ml}(t)-v^w_{mk,l}(t)\abs{q_{kl}}^2$, where $\zhat_{ml}(t)\triangleq \sum_{k=1}^{K} \what_{mk,l}(t)q_{kl}$ and $v^z_{ml}(t)\triangleq\sum_{k=1}^{K} v^w_{mk,l}(t)\abs{q_{kl}}^2$.
Then we can approximate the message \eqref{message: f-w} as
\begin{align}
&\Delta_{l \to mk}^t (w_{mk})\nonumber\\
&\propto \int \der z_{ml,k}\CN(y_{ml};z_{ml,k}+w_{mk}q_{kl},\tau_n) \nonumber \\
&\times \CN\left( z_{ml,k};\zhat_{ml}(t)-\what_{mk,l}(t)q_{kl},v^z_{ml}(t)-v^w_{mk,l}(t)\abs{q_{kl}}^2\right)\nonumber\\
&=\CN\left(y_{ml};\zhat_{ml}(t)+(w_{mk}-\what_{mk,l}(t))q_{kl},\right. \nonumber \\
&\left. \tau_n+v^z_{ml}(t)+(\abs{w_{mk}}^2-v^w_{mk,l}(t))\abs{q_{kl}}^2\right).
\end{align}
By exploiting the arguments in \cite[Eqs. (A.6)--(A.16)]{8618402}, we obtain the following approximations
        \begin{align}
\Delta_{l\to mk}^{t+1}(w_{mk})&= \CN\left(w_{mk};\what_{mk,l}(t+1),v^w_{mk,l}(t+1)\right),\\ 
\Delta_{w_{mk}}^{t+1}(w_{mk})&= \CN\left(w_{mk};\what_{mk}(t+1),v^w_{mk}(t+1)\right),\\
\prod_{l=1}^L\Delta_{l\to mk}^t (w_{mk})&=\CN(w_{mk};\ehat_{mk}(t),v^e_{mk}(t)),\label{temp142}
        \end{align}
where
\begin{subequations}\label{temp141}
        \begin{align}
        \what_{mk,l}(t+1)&\approx\what_{mk}(t+1)-v^w_{mk}(t+1)q_{kl}\gammahat_{ml}(t), \\
        v^w_{mk,l}(t+1)&\approx v^w_{mk}(t+1),\\
        v^w_{mk}(t+1)&=\frac{v^p_{mk}(t)v^e_{mk}(t)}{v^p_{mk}(t)+v^e_{mk}(t)}, \label{update: w1}\\
        \what_{mk}(t+1)&=\frac{v^p_{mk}(t)\ehat_{mk}(t)+\phat_{mk}(t)v^e_{mk}(t)}{v^p_{mk}(t)+v^e_{mk}(t)}, \label{what} \\
        v^e_{mk}(t)&=\left(\sum_{l=1}^L  v^\gamma_{ml}(t)\abs{q_{kl}}^2\right) ^{-1},\\
        \ehat_{mk}(t)&=\what_{mk}(t)+v^e_{mk}(t)\sum_{l=1}^L q_{kl}^* \gammahat_{mt}(t). \label{update: w2}
        \end{align}
        \end{subequations}
In the above equations \eqref{temp141}, we define the following variables
\begin{subequations} \label{update: w3}
        \begin{align}
        v^\gamma_{ml}(t) &= \left(v^\beta_{ml}(t)+\tau_n \right)^{-1}, \\
        \gammahat_{ml}(t) &= v^\gamma_{ml}(t)\left( y_{ml}-\betahat_{ml}(t)\right), \label{gammah} \\
        v^\beta_{ml}(t)&=\sum_{k=1}^{K}  v_{mk}^w(t)\abs{q_{kl}}^2,\\
        \betahat_{ml}(t)&=\sum_{k=1}^{K}  \what_{mk}(t)q_{kl}-v^\beta_{ml}(t)\gammahat_{ml}(i-1),\label{betah}.
        \end{align}
\end{subequations}

\subsection{Approximated messages for \eqref{message: f-g}--\eqref{message: g-f} and \eqref{message: g}} Substituting  \eqref{temp142} into \eqref{message: f-g} and \eqref{message: f-x}, we then arrive at
\begin{subequations} \label{message: f-g&f-x}
\begin{align}
        &\Delta_{k \to mn}^t(g_{mn}) \propto  \nonumber \\
        & \int \CN\left( \sum_{n=1}^{N}g_{mn}x_{nk};\ehat_{mk}(t),v^e_{mk}(t)\right)p(y_{ml}|w_{mk},\forall k)\der y_{ml} \nonumber \\
        & \times \prod_{n=1}^{N}  \Delta_{m \gets nk}^t(x_{nk})\der x_{nk} \prod_{j \neq n} \Delta_{k\gets mj}^t(g_{mj})\der g_{mj}, \\
        &\Delta_{m \to nk}^t(x_{nk}) \propto  \nonumber \\
        & \int \CN\left( \sum_{n=1}^{N}g_{mn}x_{nk};\ehat_{mk}(t),v^e_{mk}(t)\right)p(y_{ml}|w_{mk},\forall k)\der y_{ml} \nonumber \\
        &\times \prod_{n=1}^{N}\Delta_{k \to mn}^t(g_{mn})\der g_{mn} \prod_{j \neq n} \Delta_{m \gets jk}^t(x_{jk})\der x_{jk}.
\end{align}
\end{subequations}
Note that the forms of \eqref{message: f-g&f-x} have the same pattern with of \cite[Eq. (13)]{6898015}. Hence, we use the same arguments in \cite[Sec. II-D--Sec. II-E]{6898015}, and approximate that
\begin{subequations} \label{message: x&f-g}
\begin{align}
&\Delta_{x_{nk}}^{t+1}(x_{nk}) \approx p(x_{nk})\CN(x_{nk};\bhat_{nk}(t),v^b_{nk}(t)), \label{message: update x}\\
&\prod_{k=1}^{K} \Delta_{k \to mn}^t(g_{mn})=\CN(g_{mn};\chat_{mn}(t),v^c_{mn}(t)), \label{temp144}
\end{align}
\end{subequations}
where
\begin{subequations} \label{temp142}
        \begin{align}
                v^b_{nk}(t)=&\left(\sum_{m=1}^{M} \abs{\what_{mn}(t)}^2v^o_{mk}(t)\right)^{-1},\label{vb}\\
        \bhat_{nk}(t)=& \left(1-v^b_{nk}(t)\sum_{m=1}^{M} v^g_{mn}(t)v^o_{mk}(t)\right)\ghat_{nk}(t)\nonumber \\
        &+v^b_{nk}(t)\sum_{m=1}^{M} \what_{mn}^{*}(t)\ohat_{mk}(t),\label{bh}\\
        v^c_{mn}(t)=&\left(\sum_{k=1}^{K} \abs{\ghat_{nk}(t)}^2v^o_{mk}(t)\right)^{-1},\label{vc}\\
        \chat_{mn}(t)=&\left(1-v^c_{mn}(t)\sum_{k=1}^{K} v^x_{nk}(t)v^o_{mk}(t)\right)\what_{mn}(t)\nonumber \\
        &+v^c_{mn}(t)\sum_{k=1}^{K}\ghat_{nk}^{*}(t)\ohat_{mk}(t).\label{ch}
        \end{align}
\end{subequations}
In the above equations \eqref{temp142}, we define the following auxiliary variables
\begin{subequations}\label{temp57}
        \begin{align}
v^o_{mk}(t)&=\frac{v^p_{mk}(t)-v^w_{mk}(t)}{(v^p_{mk}(t))^2}, \\
        \ohat_{mk}(t)&=\frac{\what_{ml}(t)-\phat_{ml}(t)}{v^p_{mk}(t)}.
        \end{align}
\end{subequations}
Substituting \eqref{prior: X} into \eqref{message: update x}, we can obtain the mean and variance of $\Delta_{x_{nk}}^{t+1}$ as follows
\begin{subequations}\label{update: x}
\begin{eqnarray}
\xhat_{nk}(t+1)&=&\int x_{nk}\Delta_{x_{nk}}^{t+1}(x_{nk})\mathrm{d}{x_{nk}},\label{ghat}\\
v^x_{nk}(t+1)&=&\int x^2_{nk}\Delta_{x_{nk}}^{t+1}(x_{nk})\mathrm{d}{x_{nk}}-\abs{\xhat_{nk}(t+1)}^2.\label{vu}
\end{eqnarray}
\end{subequations}
By using the same arguments in \cite[Sec. II-F]{6898015}, we can rewrite \eqref{p} as
\begin{subequations}\label{update: p}
        \begin{align}
        v^p_{mk}(t)=& \nonumber \\
        \sum_{n=1}^{N} (&\abs{\ghat_{mn}(t)}^2v^x_{nk}(t)+v^g_{mn}(t) \abs{\xhat_{nk}(t)}^2+v^g_{mn}(t)v^x_{nk}(t)) ,  \label{vp}\\
        \phat_{mk}(t)=&\sum_{n=1}^{N} \xhat_{mn}(t)\ghat_{nk}(t) \nonumber \\
        -\ohat_{mk}&(i-1)\sum_{n=1}^{N}\left(\abs{\xhat_{mk}(t)}^2v^x_{nk}(t)+v^g_{mn}(t) \abs{\ghat_{nk}(t)}^2\right). \label{ph}
        \end{align}
\end{subequations}

\subsection{Approximated messages for \eqref{message: pg}--\eqref{message: g} and \eqref{message: s}} Substituting \eqref{temp144} into \eqref{message: f-s}, we obtain
\begin{align} \label{message new f-s}
& \Delta_{mn\to m'n'}^t(s_{m' n'})\propto \nonumber \\
& \int\CN\left(\sum_{m'=1}^{M'}\sum_{n'=1}^{N'}a_{B,mm'}s_{m' n'}a_{R,n' n};\chat_{mn}(t),v^c_{mn}(t)\right)\nonumber \\
& \times \prod_{(i,j)\neq (m',n')}\Delta_{mn \gets ij}^{i}(s_{ij})\der s_{ij}.
\end{align}
To simplify, we define $\textit{g}_{mn,m'n'}\triangleq \sum_{(i,j)\neq (m',n')}a_{B,mi}s_{ij}a_{R,jn}$. According to the CLT, we obtain that $\textit{g}_{mn,m' n'}$ becomes a Gaussian random variable with mean $\ghatit_{mn}(t)-a_{B,mm'}\shat_{m' n',mn}(t)a_{R,n' n}$ and variance $v^\textit{g}_{mn}(t)-a_{B,mm'}v^s_{m' n',mn}(t)a_{R,n' n}$, where
\begin{subequations}\label{g}
        \begin{align}
        \ghatit_{mn}(t)&=\sum_{m'=1}^{M'}\sum_{n'=1}^{N'}a_{B,mm'}\shat_{m' n',mn}(t)a_{R,n' n}, \\
        v^\textit{g}_{mn}(t)&=\sum_{m'=1}^{M'}\sum_{n'=1}^{N'} \abs{a_{B,mm'}}^2v^s_{m' n',mn}(t)\abs{a_{R,n' n}}^2.
        \end{align}
\end{subequations}
Substituting \eqref{g} into \eqref{message new f-s}, we obtain that
\begin{align}
&\Delta_{mn \to m'n'}^t(s_{m' n'})\nonumber \\
&\propto \CN\left(a_{B,mm'}s_{m' n'}a_{R,n' n} +\textit{g}_{mn,m' n'};\chat_{mn}(t),v^c_{mn}(t)\right)\nonumber\\
&\propto\CN\left(\ghatit_{mn}(t)-a_{B,mm'}\shat_{m' n',mn}a_{R,n' n};\right. .
\end{align}
Using the same arguments in \cite[Eqs. (A.6)--(A.16)]{8618402}, we arrive at
\begin{align}
&\prod_{m'=1}^{M'}\prod_{n'=1}^{N'}\Delta^t_{mn \gets m'n'}(s_{m' n'})=\CN\left( \textit{g}_{m l};\ghatit_{mn}(t),v^\textit{g}_{mn}(t)\right) ,\label{temp89}\\
&\prod_{m=1}^{M}\prod_{n'=1}^{N'}  \Delta_{mn \to m'n'}^t(s_{m' n'})=\CN\left( s_{m' n'};\dhat_{m' n'}(t),v^d_{m' n'}(t)\right),\label{temp90}
\end{align}
where
\begin{subequations}
        \begin{align}
        v^\textit{g}_{mn}(t)&=\sum_{m'=1}^{M'}\sum_{n'=1}^{N'} \abs{a_{B,mm'}}^2v^s_{m' n'}(t)\abs{a_{R,n' n}}^2,\label{update: g1}\\
        \ghatit_{mn}(t) &=\nonumber \\
        \sum_{m'=1}^{M'}&\sum_{n'=1}^{N'}a_{B,mm'}\shat_{m' n'}(t)a_{R,n' n}-v^\textit{g}_{mn}(t)\alphahat_{mn}(i-1),\label{muh}\\
        v^\alpha_{mn}(t)&=\left( {v^\textit{g}_{mn}(t)+v^c_{mn}(t)}\right) ^{-1}, \label{update: s1}\\
        \alphahat_{mn}(t)&=v^\alpha_{mn}(t)\left(\chat_{mn}(t)-\ghatit_{mn}(t)\right) , \label{alphah}\\
        v^d_{m' n'}(t)&=\left(\sum_{m=1}^{M}\sum_{n=1}^{N'} \abs{a_{B,mm'}}^2v^\alpha_{mn}(t)\abs{a_{R,n' n}}^2\right) ^{-1}, \\
        \dhat_{m' n'}(t)&=\shat_{m' n'} \nonumber \\
        &+v^d_{m' n'}(t)\sum_{m=1}^{M}\sum_{n=1}^{N}|a_{R,B,mm'}|^2\alphahat_{mn}(t)|a_{R,n' n}|^2 . \label{dh}
        \end{align}
\end{subequations}
Substituting \eqref{temp89} into \eqref{message: pg} along with \eqref{temp144}, we obtain
\begin{subequations}\label{update: g}
        \begin{align}
        v^g_{mn}(t+1)&=\frac{v^\textit{g}_{mn}(t)v^c_{mn}(t)}{v^\textit{g}_{mn}(t)+v^c_{mn}(t)},\\
        \ghat_{mn}(t+1)&=\frac{v^\textit{g}_{mn}(t)\chat_{mn}(t)+v^c_{mn}(t)\ghatit_{mn}(t)+v^c_{mn}(t)}{v^\textit{g}_{mn}(t)+v^c_{mn}(t)}.\label{wh}
        \end{align}
\end{subequations}
Substituting \eqref{temp90} into \eqref{message: s}, we obtain that
\begin{subequations}\label{update: s}
        \begin{align}
        \Delta_{s_{m' n'}}^{t+1}(s_{m' n'})&\propto  p(s_{m' n'})\CN\left( s_{m' n'};\dhat_{m' n'},v^d_{m' n'}\right)\\
        \shat_{m' n'}(t+1) &=\int s_{m' n'}\Delta_{s_{m' n'}}^{t+1}(s_{m' n'})\der {s_{m' n'}},\label{rh} \\
        v^s_{m' n'}(t+1) &= \nonumber \\
        \int s^2_{m' n'}&\Delta_{s_{m' n'}}^{t+1}(s_{m' n'})\der {s_{m' n'}}-\abs{\shat_{m' n'}(t+1)}^2.\label{vr}
        \end{align}
\end{subequations}

\bibliographystyle{IEEEtran}
\bibliography{reference}
\end{document}